\documentclass[11pt,a4paper]{article}
\usepackage[english]{babel}
\usepackage{amsmath,amsthm,amssymb,epsfig,latexsym}
\usepackage{color}
\usepackage{ulem}
\usepackage[numbers,square,sort&compress]{natbib}

\newcommand{\so}{\scriptscriptstyle \rm I}
\newcommand{\st}{\scriptscriptstyle \rm I\hspace{-1pt}I}
\newcommand{\sth}{\scriptscriptstyle \rm I\hspace{-1pt}I\hspace{-1pt}I}
\newcommand{\qo}{\rm i}
\newcommand{\qt}{\rm ii}

\newcommand{\be}[1]{\begin{equation}\label{#1}}
\newcommand{\ba}[1]{\begin{multline}\label{#1}}
\newcommand{\ee}{\end{equation}}
\newcommand{\ea}{\end{eqnarray}}

\newcommand{\tr}{\mathop{\rm tr}}
\newcommand{\qdet}{\mathop{\rm qdet}}

\newcommand{\bu}{\bar u}
\newcommand{\bv}{\bar v}

\newcommand{\bx}{\bar x}
\newcommand{\by}{\bar y}

\newcommand{\bxi}{\bar{\xi}}
\newcommand{\bet}{\bar{\eta}}

\newtheorem{prop}{Proposition}[section]

\newtheorem{cor}{Corollary}[section]
 \makeatletter
 \@addtoreset{equation}{section}
 \makeatother

\begin{document}


\vspace{12pt}

\begin{center}
\begin{LARGE}
{\bf On Bethe vectors in $\mathfrak{gl}_3$-invariant integrable models}
\end{LARGE}

\vspace{40pt}

\begin{large}
{A.~Liashyk${}^{a,b,c}$,
N.~A.~Slavnov${}^d$\  \footnote{
a.liashyk@gmail.com, nslavnov@mi.ras.ru}}
\end{large}

 \vspace{12mm}

${}^a$ {\it Bogoliubov Institute for Theoretical Physics, NAS of Ukraine,  Kiev, Ukraine}

\vspace{4mm}

${}^b$ {\it National Research University Higher School of Economics, Faculty of Mathematics, Moscow, Russia}

\vspace{4mm}

${}^c$ {\it Skolkovo Institute of Science and Technology, Moscow, Russia}

\vspace{4mm}

${}^d$  {\it Steklov Mathematical Institute of Russian Academy of Sciences, Moscow, Russia}

\end{center}

\vspace{4mm}


\begin{abstract}
We consider quantum integrable models solvable by the nested algebraic Bethe ansatz and possessing
$\mathfrak{gl}_3$-invariant $R$-matrix. We study a new recently proposed approach to construct on-shell
Bethe vectors of these models. We prove that the vectors constructed by this method are semi-on-shell
Bethe vectors for arbitrary values of Bethe parameters. They thus do become on-shell vectors provided
the system of Bethe equations is fulfilled.
\end{abstract}

\vspace{1cm}

\section{Introduction\label{S-I}}

Recently, a new method to construct Bethe vectors in  $\mathfrak{gl}_N$-invariant quantum spin chains  was proposed
in \cite{GroLMS17}. In the present paper we study this method by the nested algebraic Bethe ansatz (NABA) in the case of quantum integrable models with
$\mathfrak{gl}_3$-invariant $R$-matrix.

There exist several ways to study quantum integrable models with a high rank of symmetry. A nested version of
the Bethe ansatz \cite{Bet31} was proposed in \cite{Yang67,Sath68,Sath75}. In the context of the Quantum Inverse Scattering
Method (QISM) \cite{FadST79,FadT79,BogIK93L,FadLH96}, an algebraic version of this method (NABA) was developed in
\cite{Kul81,KulR82,KulR83}. One more approach based on the qKZ equation and Jackson integrals was proposed in \cite{TarVar93,TarV96,TarVar98}
and generalized to the superalgebras  in \cite{BelRag08}. We should also mention a method to construct Bethe vectors via certain projection of Drinfel'd
currents, that was developed in a series of works \cite{KhoPak05,KhoPakT07,KhoPak08,FraKhoPR08,HutLPRS17a}. The Separation of Variables (SoV) method
\cite{Skl90,Skl91} was applied to the study of $\mathfrak{gl}_3$-invariant quantum spin chains in \cite{Skl95}.

The main task of the methods listed above is to construct the eigenfunctions of the quantum Hamiltonians. Traditionally they are called
{\it on-shell Bethe vectors}. In distinction
of the $\mathfrak{gl}_2$ based models, a form of these eigenfunctions for the models with higher rank of symmetry is
quite involved. This is due to the fact that these models describe physical systems with several types of particles. Respectively,
one has to consider several creation operators corresponding to each type of excitations.

For instance, within the framework of QISM, we deal with a quantum monodromy matrix $T(u)$, whose trace plays the role of generating
functional of the integrals of motion. The upper-triangular entries of the monodromy matrix $T_{ij}(u)$ with $i<j$ are creation operators, and
a physical space of states can be generated by successive action of these operators on a referent state $|0\rangle$. In the case of
the $\mathfrak{gl}_2$ based models, there exits only one creation operator $T_{12}(u)$. Respectively,  the eigenvectors of the
quantum Hamiltonians have the form of products of these operators acting onto a referent state $|0\rangle$. However, already in the case of
the $\mathfrak{gl}_3$ based models, we deal with three creation operators, and the form of on-shell Bethe vectors immediately becomes
much more complex (see e.g. \cite{BelPRS12c} and \eqref{GBV} for explicit formulas).

It was observed in \cite{GroLMS17} that an operator used for constructing the SoV basis of the $\mathfrak{gl}_2$-invariant spin chain
can be also used for generating the basis of the on-shell Bethe vectors.  It was conjectured in \cite{GroLMS17}  that a similar effect
might take place in the spin chains with higher rank of symmetry. In particular,
in the $\mathfrak{gl}_3$-invariant spin chain one should consider an operator\footnote{%
In \cite{GroLMS17} this operator was denoted as $B^{\text{good}}(u)$. We find this notation too heavy and reduce it to $B^g(u)$.}
\begin{multline}\label{Bg}
 B^{g}(u) =  T_{23}(u)T_{12}(u-i)T_{23}(u)-T_{23}(u)T_{22}(u-i)T_{13}(u) \\
 + T_{13}(u)T_{11}(u-i)T_{23}(u)-T_{13}(u)T_{21}(u-i)T_{13}(u)
\end{multline}
for constructing the SoV basis \cite{Skl95}. Here $T_{ij}(u)$ are entries of a twisted monodromy matrix (see section~\ref{S-SM} for more details).
Then, in complete analogy with the case of $\mathfrak{gl}_2$ based models, on-shell Bethe vectors can be presented as a successive action
of $B^g(u_i)$ onto the referent state
\be{MABg}
 B^{g}(u_1)\dots  B^{g}(u_a)|0\rangle.
 \ee
This conjecture was justified by the computer calculation, however, an analytical proof is lacking so far.
The goal of this paper is to find such the proof.

Our proof of representation \eqref{MABg} is given within the framework of NABA. We show that representation \eqref{MABg} for
on-shell Bethe vectors holds not only for spin chains, but for a more wide class of integrable models possessing
$\mathfrak{gl}_3$-invariant $R$-matrix. In particular, we do not use the SoV method.

The paper is organized as follows.  We recall basic notions of NABA in section~\ref{S-BNNABA}. There we
also give a standard description of Bethe vectors within this method.  Section~\ref{S-SM} is devoted to
special NABA-solvable models that usually are applied to the systems of physical interest. The main
results of our paper are gathered in  section~\ref{S-MR}. There we give explicit representation of the
states \eqref{MABg} in terms of the monodromy matrix entries acting on the pseudovacuum  vector. We also
describe a relationship between the states \eqref{MABg} and the Bethe vectors obtained by the standard
NABA approach. In the rest of the paper we give the proofs of the results of section~\ref{S-MR}.
We identify the state \eqref{MABg} with a Bethe vector in section~\ref{S-PC}. In
section~\ref{S-ABgBV} we compute the action of the operator $B^g(u)$ on a generic Bethe vector. Finally,
in section~\ref{S-PP} we express the state \eqref{MABg} in terms  of the monodromy matrix entries acting on the pseudovacuum  vector.
Several auxiliary identities for rational functions are gathered in appendix~\ref{A-DWPF}.
Appendix~\ref{S-PCBTTBV} contains a proof of connection between two types of Bethe vectors considered in
the paper.  Finally, the formulas of the action of the monodromy matrix entries onto the Bethe vectors
are given in appendix~\ref{S-AF}.

\section{Basic notions of NABA \label{S-BNNABA}}

We consider quantum integrable models solvable by NABA and possessing the $\mathfrak{gl}_3$-invariant $R$-matrix
\be{Rmat}
R(u,v)=\mathbf{I}\otimes\mathbf{I}+g(u,v)\mathbf{P}, \qquad g(u,v)=\frac{c}{u-v}.
\ee
Here $\mathbf{I}$ is the identity matrix in $\mathbf{C}^3$, $\mathbf{P}$ is the permutation matrix
in $\mathbf{C}^3\otimes \mathbf{C}^3$, and $c$ is a constant\footnote{To compare our presentation
with the results of \cite{GroLMS17} one should set $c=i$.}.

The monodromy matrix $T(u)$ is a $3\times 3$ matrix with operator-valued entries $T_{ij}(u)$ acting in a Hilbert space
$\mathcal{H}$. Their commutation relations are
give by an $RTT$-relation
\be{RTT}
R(u,v)\bigl(T(u)\otimes \mathbf{I}\bigr)\bigl(\mathbf{I}\otimes T(v)\bigr) = \bigl(\mathbf{I}\otimes T(v)\bigr)\bigl(T(u)\otimes \mathbf{I} \bigr) R(u,v).
\ee
It follows from \eqref{RTT} that an operator
\be{trT}
\mathcal{T}(u)=\tr T(u)=\sum_{i=1}^3 T_{ii}(u)
\ee
has the following property: $[\mathcal{T}(u),\mathcal{T}(v)]=0$ for arbitrary $u$ and $v$. This operator is called a transfer matrix. It plays the role
of a generating functional of the integrals of motion of a quantum model under consideration. One of the main tasks of NABA is to find
eigenvectors of this operator.

If a $3\times 3$ $c$-number matrix $K$ is such that $[R(x,y),K\otimes K]=0$, then the matrices $KT(u)$ and
$T(u)K$ also satisfy the $RTT$-relation \eqref{RTT}. A peculiarity of the $R$-matrix \eqref{Rmat} is that
$[R(x,y),K\otimes K]=0$ holds for arbitrary $K\in \mathfrak{gl}_3$. In particular, if $K$ is invertible, then one can
consider a transformation $T(u)\to T^{(K)}(u)=KT(u)K^{-1}$. Obviously, this transformation preserves the transfer matrix.

Besides the monodromy matrix $T(u)$, we also will consider a matrix $\widehat{T}(u)$ that is closely associated to
a quantum comatrix \cite{MolNO96,Mol07}. First, we introduce quantum minors
\be{Sklmin}
t^{j_1,j_2}_{k_1,k_2}(u)=T_{j_1,k_1}(u)T_{j_2,k_2}(u-c)-T_{j_2,k_1}(u)T_{j_1,k_2}(u-c).
\ee
The  entries of the quantum comatrix $\widetilde{T}_{jk}(u)$ then are given by
\be{SklAldop}
\widetilde{T}_{jk}(u)=(-1)^{j+k}t_{\bar j}^{\bar k}(u),
\ee
where $\bar j=\{1,2,3\}\setminus j$. The  quantum comatrix plays the role of the inverse monodromy
matrix due to
\be{comat}
\widetilde{T}(u-c) T(u)=\qdet T(u)\;\mathbf{I},
\ee
where $\qdet T(u)$ is a quantum determinant of $T(u)$ \cite{IzeK81,KulS82,MolNO96,Mol07}.

The matrix $\widehat{T}(u)$ is defined as the transposition of $\widetilde{T}(u)$ with respect to
the secondary diagonal:
\be{Qmin}
\widehat{T}_{jk}(u)=\widetilde{T}_{4-k,4-j}(u).
\ee
It is known \cite{NazT98,MolNO96,BelPRS12c,Mol07}  that a mapping
$\phi: T(u)\mapsto \widehat{T}(u)$ is an automorphism of the $RTT$-algebra \eqref{RTT}.
Thus, the matrix $\widehat{T}(u)$ satisfies the $RTT$-relation with the same $R$-matrix \eqref{Rmat}.

Using the matrix $\widehat{T}(u)$ we can write down the operator $B^g(u)$ \eqref{Bg} in a more compact form
\be{Bg-ND}
B^g(u)=T_{23}(u)\widehat{T}_{13}(u)-T_{13}(u)\widehat{T}_{12}(u).
\ee
Similar representation for $B^g$ was used in \cite{Skl95}.
%
%

\subsection{Notation}
Besides the function $g(u,v)$ we also introduce two new functions
\be{univ-not}
 f(u,v)=1+g(u,v)=\frac{u-v+c}{u-v},\qquad h(u,v)=\frac{f(u,v)}{g(u,v)}=\frac{u-v+c}{c}.
\ee
The following obvious properties of the functions introduced above are useful:
 \be{propert}
 g(u,v)=-g(v,u),\quad h(u-c,v)=\frac1{g(u,v)},\quad g(u+c,v)=\frac1{h(u,v)},\quad  f(u-c,v)=\frac1{f(v,u)}.
 \ee

Before giving a description of the Bethe vectors we formulate a convention on the notation.
We denote sets of variables by a bar:  $\bu$, $\bv$, and so on.
Individual elements
of the sets are denoted by subscripts:  $u_j$, $v_k$,  and so on. Notation $\bar u+c$ means that the
constant $c$ is added to all the elements of the set $\bar u$.
Subsets of variables are denoted by roman indices: $\bu_{\so}$, $\bu_{\st}$, $\bv_{\qt}$, and so on.
In particular, we consider partitions of sets into subsets. Then the notation $\bu\Rightarrow\{\bu_{\so},\;\bu_{\st}\}$ means that the
set $\bu$ is divided into two disjoint subsets $\bu_{\so}$ and $\bu_{\st}$. The order of the elements in each subset is not essential.
A special notation $\bu_j$ is used
for subsets complementary to the element $u_j$, that is, $\bu_j=\bu\setminus u_j$, $\bv_k=\bv\setminus v_k$ and so on.

In order to avoid too cumbersome formulas we use shorthand notation for products of
functions depending on one or two variables. Namely, if the functions $g$, $f$, and $h$  depend on sets of variables, this means that
one should take the product over the corresponding set.
For example,
 \be{SH-prod}
h(\bu,v)=\prod_{u_j\in\bu} h(u_j,v);\quad
 g(z_i, \bar z_i)= \prod_{\substack{z_j\in\bar z\\ z_j\ne z_i}} g(z_i, z_j);\quad
 f(\bu_{\st},\bu_{\so})=\prod_{u_j\in\bu_{\st}}\prod_{u_k\in\bu_{\so}} f(u_j,u_k).
 \ee
In the last equation of \eqref{SH-prod} the set $\bar u$ is divided into two subsets
$\bu_{\so}$,\; $\bu_{\st}$, and the double product is taken with respect to all
$u_k$ belonging to $\bu_{\so}$ and all $u_j$ belonging to $\bu_{\st}$.
We use the same prescription for the products of commuting operators and their vacuum eigenvalues $\lambda_i$  (see \eqref{Tjj})
 \be{SH-prod1}
\lambda_i(\bu)=\prod_{u_j\in\bu} \lambda_i(u_j);\quad
 T_{ij}(\bv_{\so})= \prod_{v_k\in\bv_{\so}} T_{ij}(v_k).
 \ee
By definition, any product over the empty set is equal to $1$. A double product is equal to $1$ if at least one of the sets
is empty.

\subsection{Bethe vectors}

Now we pass to the description of Bethe vectors. They belong to a Hilbert space $\mathcal{H}$, in which
the operators $T_{ij}(u)$ act. We assume that this space contains a referent state (pseudovacuum vector) $|0\rangle$ such that
 \be{Tjj}
 \begin{aligned}
 &T_{jj}(u)|0\rangle=\lambda_j(u)|0\rangle,\\
 &T_{ij}(u)|0\rangle=0, \qquad i>j,
 \end{aligned}
 \ee
where $\lambda_j(u)$ are some scalar functions. Generically, they are free functional parameters.

The action of the operators $T_{ij}(u)$ with $i<j$ onto pseudovacuum  is free. Within the framework of NABA, it is assumed that
successive action of these operators onto $|0\rangle$ generates vectors of the space $\mathcal{H}$. Bethe vectors are
special polynomials in $T_{ij}(u)$ with $i<j$ acting on $|0\rangle$. Their explicit form will be given  later. Here we would like to mention
that in the models with $\mathfrak{gl}_3$-invariant $R$-matrix Bethe vectors depend on two sets of complex parameters
$\bu=\{u_1,\dots,u_a\}$ and $\bv=\{v_1,\dots,v_b\}$ called Bethe parameters. We denote these vectors by $\mathbb{B}_{a,b}(\bu;\bv)$,
where $a$ and $b$ respectively
are the cardinalities of the sets $\bu$ and $\bv$. A characteristic property of
the Bethe vectors is that they become eigenvectors of the
transfer matrix $\mathcal{T}(z)=\tr T(z)$ provided $\bu$ and $\bv$ enjoy ceratin constraint. In this case they are called  on-shell Bethe vectors.
Otherwise, if $\bu$ and $\bv$ are generic complex numbers, then the corresponding vector is called {\it off-shell Bethe vector}.

In physical models, vectors of the space $\mathcal{H}$  describe states with quasiparticles (excitations) of two different types (colors).
We say that a state has coloring $\{a,b\}$, if it contains $a$ quasiparticles of the color $1$ and $b$ quasiparticles of the color $2$.
The vector $|0\rangle$ has zero coloring.  The operator $T_{12}$ is the creation operator of quasiparticles of the first color,
while the operator $T_{23}$  creates  quasiparticles of the second color. The operator $T_{13}$ creates one quasiparticle of the first color and
one quasiparticle of the second color. The diagonal operators $T_{ii}$ are neutral, the matrix elements $T_{ij}$ with $i>j$ play the role of
annihilation operators.  Generally, there are no restrictions on
the coloring $\{a,b\}$, thus, the parameters $a$ and $b$ are arbitrary non-negative integers. In specific models, some restrictions
may appear.

Different methods to construct Bethe vectors were developed in \cite{KulR83,TarVar98,BelRag08,KhoPak05}. Several equivalent explicit representations
were found in \cite{BelPRS12c}. One of this representations reads
\be{GBV}
\mathbb{B}_{a,b}(\bu;\bv)=\sum_{n=0}^{\min(a,b)}\sum_{\#\bu_{\so}=\#\bv_{\so}=n} \frac{K_n(\bv_{\so}|\bu_{\so})f(\bu_{\so},\bu_{\st})
f(\bv_{\st},\bv_{\so})}{\lambda_2(\bv_{\st})\lambda_2(\bu)g(\bv,\bu)}
T_{13}(\bu_{\so}) T_{12}(\bu_{\st})T_{23}(\bv_{\st})|0\rangle.
\ee
Recall that here we use the shorthand notation \eqref{SH-prod}, \eqref{SH-prod1} for the products of the operators $T_{ij}$ and the functions
$\lambda_2$, $f$, and $g$.
The sum in \eqref{GBV} is taken over partitions of the sets
$\bu\Rightarrow\{\bu_{\so},\bu_{\st}\}$  and $\bv\Rightarrow\{\bv_{\so},\bv_{\st}\}$ such that
$\#\bu_{\so}=\#\bv_{\so}=n$, where $n=0,1,\dots,\min(a,b)$. It is easy to see that each term of this sum has
a fixed coloring $\{a,b\}$, and thus, Bethe vector $\mathbb{B}_{a,b}(\bu;\bv)$ has coloring that coincides with the cardinalities
of the Bethe parameters. We would like to stress
that generically there is no any restriction on the cardinalities
of the Bethe parameters $\bu$ and $\bv$. In particular, one might have $a<b$, that is $\#\bu<\#\bv$.

The function $K_n(\bv_{\so}|\bu_{\so})$ in \eqref{GBV} is a partition function of the six-vertex model with domain wall boundary condition (DWPF)
\cite{Kor82,Ize87}. It depends on two sets of variables $\bar v$ and $\bar u$; the subscript shows that
$\#\bar v=\#\bar u=n$. The function $K_n$ has the following determinant representation \cite{Ize87}:
\begin{equation}\label{K-def}
K_n(\bar v|\bar u)
=h(\bar v,\bar u)\left( \prod_{j<k}^n g(v_j,v_k)g(u_k,u_j)\right)
\det_n \left(\frac{g(v_j,u_k)}{h(v_j,u_k)}\right).
\end{equation}
Some properties of $K_n$  are gathered in appendix~\ref{A-DWPF}.

Observe  that the normalization in \eqref{GBV} differs from the normalization of Bethe vectors used in
\cite{BelPRS12c}. The present normalization is chosen so that the Bethe vector does not have singularities for
$v_j=u_k$ and $v_j-c=u_k$.

We also consider Bethe vectors $\widehat{\mathbb{B}}_{a,b}(\bu;\bv)$ which correspond to the monodromy matrix $\widehat{T}(u)$. They
have the form
\be{hGBV}
\widehat{\mathbb{B}}_{a,b}(\bu;\bv)=\sum_{n=0}^{\min(a,b)}\sum_{\#\bu_{\so}=\#\bv_{\so}=n} \frac{K_n(\bv_{\so}|\bu_{\so})f(\bu_{\so},\bu_{\st})
f(\bv_{\st},\bv_{\so})}{\hat\lambda_2(\bv_{\st})\hat\lambda_2(\bu)g(\bv,\bu)}
\widehat{T}_{13}(\bu_{\so})\widehat{T}_{12}(\bu_{\st})\widehat{T}_{23}(\bv_{\st})|0\rangle,
\ee
where $\hat\lambda_2(z)=\lambda_1(z)\lambda_3(z-c)$.

The  automorphism $T(u)\mapsto\widehat{T}(u)$ generates a connection between the Bethe vectors $\mathbb{B}$ and $\widehat{\mathbb{B}}$: 
\be{hB-B}
\widehat{\mathbb{B}}_{b,a}(\bv+c;\bu)=(-1)^{a+b+ab}\frac{\lambda_2(\bu)\lambda_2(\bv)}{\lambda_1(\bu)\lambda_3(\bv)}\;\mathbb{B}_{a,b}(\bu;\bv).
\ee
The proof is given in appendix~\ref{S-PCBTTBV}.

 We have mentioned already that a generic Bethe vector becomes an on-shell Bethe vector,
if the parameters $\bu$ and $\bv$  satisfy a special constraint. This constraint is known as a system of Bethe equations and has
the following form:
\be{AEigenS-1}
\begin{aligned}
\frac{\lambda_1(u_j)}{\lambda_2(u_j)}&
=\frac{f(u_j,\bu_j)}{f(\bu_j,u_j)}f(\bv,u_j), \qquad j=1,\dots, a,\\
\frac{\lambda_2(v_k)}{\lambda_3(v_k)}&=\frac{f(v_k,\bv_k)}{f(\bv_k,v_k)}\frac1{f(v_k,\bu)}, \qquad k=1,\dots, b.
\end{aligned}
\ee
If the system  \eqref{AEigenS-1} is fulfilled, then
\be{Left-act}
\mathcal{T}(z)\mathbb{B}_{a,b}(\bu;\bv) = \tau(z|\bu,\bv)\,\mathbb{B}_{a,b}(\bu;\bv),
\ee
where
\be{tau-def}
\tau(z)\equiv\tau(z|\bu,\bv)=\lambda_1(z)f(\bu,z)+\lambda_2(z)f(z,\bu)f(\bv,z)+\lambda_3(z)f(z,\bv).
\ee

Below we will need the action formulas of the operators $T_{ij}(z)$ and $\widehat{T}_{ij}(z)$ on the generic
Bethe vectors. They were obtained in \cite{BelPRS12c}. We give the list of necessary formulas  in appendix~\ref{S-AF}.

\section{Special NABA-solvable models \label{S-SM}}

At the first sight, a method to construct on-shell Bethe vectors by means of the operator $B^g(u)$ \eqref{Bg} contradicts to the
content of the previous section. Indeed, according to the general scheme, the on-shell Bethe vector depends on {\it two} sets of
variables subject to the equations \eqref{AEigenS-1}. At the same time, vector \eqref{MABg} depends on only one set of variables.
The solution of this contradiction lies in the fact that in some models there is a kind of hierarchy between the variables $\bu$ and $\bv$:
the set $\bu$ plays a basic role, while the variables $\bv$ are auxiliary. In particular, the system of Bethe equations can be reformulated as a
constraint on the Bethe parameters $\bu$ only (see \eqref{BE-det} below).

This class of models includes the $XXX$ $SU(3)$-invariant Heisenberg chain, for which the operator $B^g(u)$ was originally constructed
in \cite{GroLMS17}. A   characteristic  property of these models is that only the operators $T_{12}(u)$ and $T_{13}(u)$ are
true creation operators, while $T_{23}(u)|0\rangle=0$. In spite of these models are a particular case of the models considered above,
they find a wide application in physics\footnote{%
One can also consider models, in which  $T_{12}(u)|0\rangle=0$, while $T_{23}(u)$ and $T_{13}(u)$ are true creation operators.
This case is equivalent to the one considered in this paper, due to an automorphism of the $RTT$-algebra \eqref{RTT} with respect to the replacement
$T_{ij}(u)\to T_{4-j,4-i}(-u)$.}.

Consider a monodromy matrix $T^0(u)$ such that $T^0_{23}(u)|0\rangle=0$. This condition immediately implies a restriction on the
vacuum eigenvalues $\lambda_j(u)$. Indeed, it follows from the $RTT$-relation that
\be{Com3223}
[T^0_{32}(u),T^0_{23}(v)] = g(u,v)\bigl(T^0_{22}(v)T^0_{33}(u)-T^0_{22}(u)T^0_{33}(v)\bigr).
\ee
Acting with this equation onto  $|0\rangle$ we obtain
\be{Com3223lam}
0=\bigl(\lambda_2(v)\lambda_3(u)-\lambda_2(u)\lambda_3(v)\bigr)|0\rangle,
\ee
leading to $\lambda_2(u)=\kappa \lambda_3(u)$, where $\kappa$ is a constant. Without loss of generality we can set
$\lambda_2(z)=\kappa$ and $\lambda_3(z)=1$. At the same time, the vacuum eigenvalue $\lambda_1(z)$ still remains a
free functional parameter. Below we omit the subscript and denote it $\lambda_1(z)=\lambda(z)$.

Bethe equations \eqref{AEigenS-1} take the form
\begin{subequations}
\begin{align}
&\lambda(u_j)=\kappa\frac{f(u_j,\bu_j)}{f(\bu_j,u_j)}f(\bv,u_j), &\qquad j=1,\dots, a,\label{constV}\\
&\kappa f(v_k,\bu)=\frac{f(v_k,\bv_k)}{f(\bv_k,v_k)}, &\qquad k=1,\dots, b.\label{2constV}
\end{align}
\end{subequations}
One can show (see e.g. \cite{GorZZ14}) that this system implies
\be{BE-det}
\det_a\left(\delta_{jk}+\alpha\lambda(u_j)\frac{f(\bu_j,u_j)}{h(u_k,u_j)}\right)=
(1+\alpha)^b(1+\alpha\kappa)^{a-b}.
\ee
Here $\alpha$ is a complex number. Equation \eqref{BE-det} should be valid for an arbitrary value of
this parameter. As both sides of \eqref{BE-det} are polynomials in $\alpha$ of degree $a$, this condition
is equivalent to a set of $a$ equations for $a$ variables $\bu=\{u_1,\dots,u_a\}$ (the free terms in
both sides obviously are equal to $1$). We see that the set of auxiliary variables $\bv$ is eliminated.

According to the coloring prescriptions, quasiparticles of the second color now can be created by the action
of the operator $T^0_{13}(u)$ only. Since this operator simultaneously creates a quasiparticle of the first
color, we conclude that the coloring of any state in these models has a property $b\le a$. In particular,
Bethe vectors $\mathbb{B}^0_{a,b}(\bu;\bv)$ for such the monodromy matrix possess this property. Their explicit form also simplifies:
\be{BVpartr}
\mathbb{B}^0_{a,b}(\bu;\bv)=\sum_{\#\bu_{\so}=b} \frac{K_b(\bv|\bu_{\so})f(\bu_{\so},\bu_{\st})}{\kappa^{a}g(\bv,\bu)}
T^0_{13}(\bu_{\so}) T^0_{12}(\bu_{\st})|0\rangle.
\ee
In distinction of \eqref{GBV}, here the sum is taken over partitions of the set
$\bu\Rightarrow\{\bu_{\so},\bu_{\st}\}$ such that $\#\bu_{\so}=b$, while the set $\bv$ is not divided into subsets.
We see that a generic off-shell Bethe vector $\mathbb{B}^0_{a,b}(\bu;\bv)$ still depends on the set of auxiliary
Bethe parameters $\bv$. We will show, however, that the auxiliary parameters
can be eliminated from on-shell Bethe vectors, as it was done for the system of Bethe equations.

Thus, for the models with the monodromy matrix $T^0(u)$, one can actually restrict himself with a one set of the Bethe
parameters only. However, if we substitute the operators $T^0_{ij}(u)$ into equation \eqref{Bg} for $B^g(u)$, then we see that
$B^g(u)|0\rangle=0$. This is due to the fact that $T^0_{23}(u)|0\rangle=0$. Thus, the operator \eqref{Bg} cannot be used as a creation
operator in these models.

A nontrivial action of  $B^g(u)$ onto the pseudovacuum vector can be provided by an appropriate twist transformation
\be{TwMM}
T(u)=KT^0(u)K^{-1}.
\ee
In paper \cite{GroLMS17}, a generic twist matrix $K$ was considered. We restrict ourselves with
a `minimal' twist, which provides a condition $T_{23}(u)|0\rangle\ne0$, but does not change the action of other operators
$T_{ij}$ onto $|0\rangle$. Let
\be{gf}
K=I+\frac{\beta}{1-\kappa}E_{23},
\ee
where $\beta\ne 0$ is a complex number and $E_{23}$ is an elementary unit matrix $(E_{23})_{ij}=\delta_{i2}\delta_{j3}$. It is easy to see that
the matrix $T(u)$ has the same vacuum eigenvalues $\lambda_1(z)=\lambda(z)$, $\lambda_2(z)=\kappa$,
and $\lambda_3(z)=1$. However, now we have $T_{23}(u)|0\rangle=\beta|0\rangle$ provided $\kappa\ne 1$.

Of course, the twist matrix \eqref{gf} is not the only matrix, ensuring the condition $T_{23}(u)|0\rangle\ne0$.
We discuss more general twists in Conclusion.

\section{Main results\label{S-MR}}

We are now in position to formulate our main results.


\begin{prop}\label{P-Bg-BV}
Let the vacuum eigenvalues of the monodromy matrix $T(u)$ be given be equations
\be{VacEV}
T_{11}(u)|0\rangle=\lambda(u)|0\rangle,\qquad T_{22}(u)|0\rangle=\kappa|0\rangle,\qquad
T_{33}(u)|0\rangle=|0\rangle,
\ee
and $T_{23}(u)|0\rangle=\beta|0\rangle$. Let  $\bu$ and $\bv$ be two sets of complex numbers such that $\#\bu=a$, $\#\bv=b$, and the
constraint \eqref{constV} is fulfilled. Then  Bethe vector $\mathbb{B}_{a,b}(\bu,\bv)$ has the following representation:
\begin{multline}\label{actBg}
\mathbb{B}_{a,b}(\bu,\bv)=
\sum_{n=0}^a\frac{\beta^{b-n}}{\kappa^{a+b} g(\bv,\bu)}\sum_{s=0}^n\sum_{\substack{\#\bu_{\so}=s\\\#\bu_{\st}=n-s}}
(-\kappa)^{n-s}\lambda(\bu_{\so})f(\bu_{\st},\bu_{\so})f(\bu_{\sth},\bu_{\so})f(\bu_{\st},\bu_{\sth})\\
\times T_{13}(\bu_{\so})T_{13}(\bu_{\st})T_{12}(\bu_{\sth})|0\rangle.
\end{multline}
Here the sum is taken over partitions of the set $\bu$ into three subsets $\bu\Rightarrow\{\bu_{\so},\bu_{\st},\bu_{\sth}\}$.
The cardinalities of the subsets are shown explicitly by the subscripts of the sum symbol in \eqref{actBg}.
\end{prop}

The proof of proposition~\ref{P-Bg-BV} is based on the  explicit representation
for the Bethe vectors \eqref{GBV}. This is done in section~\ref{S-PC}. Here we give several comments on this proposition.

The condition \eqref{constV} is a part of Bethe equations, therefore, the corresponding Bethe vector
can be called a {\it semi-on-shell Bethe vector} \cite{HutLPRS16c}. The constraint \eqref{constV} is a system of $a$ equations
for $a+b$ variables. In particular, if\hspace{2pt}\footnote{Recall that
due to $T_{23}(z)|0\rangle \ne 0$ we have no restriction $b\le a$.}
$b\ge a$, then we can consider \eqref{constV} as the system of equations for the parameters $v_k$, $k=1,\dots,b$. At the same time the parameters
$\bu$ remain generic complex numbers, and one can easily show that the system is solvable. Furthermore, it follows from
representation \eqref{actBg} that if $v_k$, $k=1,\dots,b$ and
$v'_k$, $k=1,\dots,b'$ are two different solutions to the system \eqref{constV}, then
\be{relBbbp}
\kappa^{b'-b}g(\bv',\bu)\mathbb{B}_{a,b'}(\bu,\bv')=\beta^{b'-b}g(\bv,\bu)\mathbb{B}_{a,b}(\bu,\bv).
\ee
This property  is due to the very specific action of the operator $T_{23}(z)$ onto the pseudovacuum vector:
$T_{23}(z)|0\rangle=\beta|0\rangle$. Thus, two  semi-on-shell Bethe vectors with different sets of the Bethe parameters $\bv$ and $\bv'$
actually are proportional to each other. In fact, for an appropriate normalization, semi-on-shell Bethe
vectors \eqref{actBg} do not depend on the parameters of the set $\bv$.

Proposition~\ref{P-Bg-BV} implies that on-shell Bethe vectors also have representation \eqref{actBg}. In this case the parameters
$\bu$ and $\bv$ enjoy the additional set of equations \eqref{2constV}. We see, however, that the
condition \eqref{constV} is already sufficient to eliminate the parameters $\bv$ from the representation for the
Bethe vector. They are only included in the normalization factor.


Now we give an explicit representation for the multiple action of the operator $B^g$ onto pseudovacuum
vector $|0\rangle$. It was shown in \cite{GroLMS17} that $[B^g(u),B^g(v)]=0$ for arbitrary $u$ and $v$. Thus, given a set
$\bu=\{u_1,\dots,u_a\}$, then the notation
\be{Not}
B^g(\bu)=\prod_{j=1}^aB^g(u_j)
\ee
is well defined.

\begin{prop}\label{P-Bg-BV0}
Let the vacuum eigenvalues of the monodromy matrix $T(u)$ be as in proposition~\ref{P-Bg-BV}
and $T_{23}(u)|0\rangle=\beta|0\rangle$. Let a set $\bu$ consist of generic complex numbers and $\#\bu=a$. Then
\begin{multline}\label{Bg-Maction}
B^g(\bu)|0\rangle=
\sum_{n=0}^a\beta^{2a-n}\sum_{s=0}^n\sum_{\substack{\#\bu_{\so}=s\\\#\bu_{\st}=n-s}}
(-\kappa)^{n-s}\lambda(\bu_{\so})f(\bu_{\st},\bu_{\so})f(\bu_{\sth},\bu_{\so})f(\bu_{\st},\bu_{\sth})\\
\times T_{13}(\bu_{\so})T_{13}(\bu_{\st})T_{12}(\bu_{\sth})|0\rangle.
\end{multline}
Here the sum  over partitions of $\bu$ is taken as in proposition~\ref{P-Bg-BV}.
\end{prop}

This proposition gives the result of multiple action of the operator $B^g$ onto $|0\rangle$ in terms of multiple actions of the
creation operators $T_{12}$ and $T_{13}$. The proof of proposition~\ref{P-Bg-BV0} is given in section~\ref{S-PP}.

Comparing  \eqref{Bg-Maction} and \eqref{actBg} we immediately arrive at

\begin{cor}\label{C-Bg-BV0}
Under the conditions of propositions~\ref{P-Bg-BV} and~\ref{P-Bg-BV0}
\be{Bg-BV}
B^g(\bu)|0\rangle=\beta^{2a-b}\kappa^{a+b} g(\bv,\bu)\mathbb{B}_{a,b}(\bu,\bv).
\ee
\end{cor}

Thus, if the Bethe parameters $\bu$ and $\bv$ satisfy Bethe equations \eqref{constV}, \eqref{2constV}, then the vector
$B^g(\bu)|0\rangle$ is on-shell Bethe vector, as it is proportional to the on-shell Bethe vector
$\mathbb{B}_{a,b}(\bu,\bv)$. One can also consider the vector $B^g(\bu)|0\rangle$ for generic complex $\bu$. Equation
\eqref{Bg-BV} remains true in this case, if the set $\bv$ satisfies the system \eqref{constV}. Due to the property
\eqref{relBbbp} one can always provide the solvability of this system for generic complex $\bu$.

\section{Proof of proposition~\ref{P-Bg-BV}\label{S-PC}}

We begin with an explicit form of Bethe vectors corresponding to the twisted monodromy matrix $T(u)$ \eqref{TwMM}. This form
follows from the general representation \eqref{GBV}, where one should take into account the condition $T_{23}(u)|0\rangle=\beta|0\rangle$.
Then
\be{BVtwT}
\mathbb{B}_{a,b}(\bu;\bv)=\sum_{n=0}^{\min(a,b)}\sum_{\#\bu_{\so}=\#\bv_{\so}=n} \frac{\beta^{b-n}K_n(\bv_{\so}|\bu_{\so})f(\bu_{\so},\bu_{\st})
f(\bv_{\st},\bv_{\so})}{\kappa^{a+b-n}g(\bv,\bu)}
T_{13}(\bu_{\so}) T_{12}(\bu_{\st})|0\rangle.
\ee
Here, like in \eqref{GBV}, the sum is taken over partitions of the sets
$\bu\Rightarrow\{\bu_{\so},\bu_{\st}\}$  and $\bv\Rightarrow\{\bv_{\so},\bv_{\st}\}$. The subscripts of the sums show that
the partitions satisfy restrictions
$\#\bu_{\so}=\#\bv_{\so}=n$, where $n=0,1,\dots,\min(a,b)$.

The sum over partitions
$\bv\Rightarrow\{\bv_{\so},\bv_{\st}\}$ can be transformed into a sum over additional partitions of the subset $\bu_{\so}$
via \eqref{ident}, in which one should set
$\bx=\bv$ and $\by=\bu_{\so}$.  Then
\be{ident0}
\sum_{\#\bv_{\so}=n} K_n(\bv_{\so}|\bu_{\so})f(\bv_{\st},\bv_{\so})=\sum_{s=0}^n\sum_{\#\bu_{\qo}=s} (-1)^{n-s}f(\bu_{\qo},\bu_{\qt})f(\bv,\bu_{\qo}).
\ee
Here in the lhs, the sum is taken over partitions $\bv\Rightarrow\{\bv_{\so},\bv_{\st}\}$ so that $\#\bv_{\so}=n$.
In the rhs, the sum is taken over all possible partitions $\bu_{\so}\Rightarrow\{\bu_{\qo},\bu_{\qt}\}$. Substituting this into
\eqref{BVtwT} we find
\begin{multline}\label{BVt-expl1}
\mathbb{B}_{a,b}(\bu,\bv)=
\sum_{n=0}^{a}\frac{\beta^{b-n}}{\kappa^{a+b-n}g(\bv,\bu)}
\sum_{s=0}^n\sum_{\substack{\#\bu_{\qo}=s\\\#\bu_{\qt}=n-s}}
(-1)^{n-s}f(\bu_{\qo},\bu_{\qt})f(\bv,\bu_{\qo})f(\bu_{\qo},\bu_{\st})f(\bu_{\qt},\bu_{\st})\\
\times T_{13}(\bu_{\qo})T_{13}(\bu_{\qt})T_{12}(\bu_{\st})|0\rangle.
\end{multline}
In \eqref{BVt-expl1}, the sum is taken over partitions of the set $\bu$ into three subsets
$\bu\Rightarrow\{\bu_{\qo},\bu_{\qt},\bu_{\st}\}$. The cardinalities of subsets are shown explicitly by the subscripts of the sum.

Note that we have replaced the upper summation limit $\min(a,b)$ with $a$ in the sum over $n$. If $a\le b$, then this replacement certainly is possible.
If $a>b$, then all the terms in the sum over $n$ with $n>b$ vanish due to proposition~\ref{P-A12}. Indeed, due to this proposition
the sum in the rhs of \eqref{ident0} gives a determinant \eqref{identA3}. The latter vanishes for $n>b$. Thus, if $a>b$, then the sum
in \eqref{BVt-expl1} actually breaks at $n=b$.

Suppose that $\mathbb{B}_{a,b}(\bu,\bv)$ is a semi-on-shell Bethe vector whose Bethe parameters satisfy the condition \eqref{constV}.
Then taking the product of equations \eqref{constV} over subset $\bu_{\qo}$ we find
\be{constVpr}
f(\bv,\bu_{\qo})=\kappa^{-s}\lambda(\bu_{\qo})
\frac{f(\bu_{\st},\bu_{\qo})f(\bu_{\qt},\bu_{\qo})}{ f(\bu_{\qo},\bu_{\qt})f(\bu_{\qo},\bu_{\st})}.
\ee
Substituting this into \eqref{BVt-expl1} we arrive at
\begin{multline}\label{BVt-expl2}
\mathbb{B}_{a,b}(\bu,\bv)=
\sum_{n=0}^a\frac{\beta^{b-n}}{\kappa^{a+b}g(\bv,\bu)}\sum_{s=0}^n\sum_{\substack{\#\bu_{\qo}=s\\\#\bu_{\qt}=n-s}}
(-\kappa)^{n-s}\lambda(\bu_{\qo})f(\bu_{\qt},\bu_{\qo})f(\bu_{\st},\bu_{\qo})f(\bu_{\qt},\bu_{\st})\\
\times T_{13}(\bu_{\qo})T_{13}(\bu_{\qt})T_{12}(\bu_{\st})|0\rangle.
\end{multline}
This representation coincides with \eqref{actBg} up to the labels of the subsets. Thus, proposition~\ref{P-Bg-BV} is proved.

\section{Action of $B^g(z)$ on Bethe vectors\label{S-ABgBV}}

We use induction over $a$ in order to prove proposition~\ref{P-Bg-BV0}. However, before doing this, we find the action
of the operator $B^g(z)$ on an arbitrary Bethe vector $\mathbb{B}_{a,b}(\bu,\bv)$. This will give us a necessary tool
for the proof.

Below, for some time, we
do not use  restrictions $T_{23}(z)|0\rangle=\beta|0\rangle$, $\lambda_2(z)=\kappa$, and $\lambda_3(z)=1$.
Instead, we consider the most general case of the monodromy matrix. In order to avoid new notation, we still
denote this monodromy matrix by $T(z)$. However, we do not assume that the action of $T_{23}(z)$ has some peculiarity,
nor do we impose any restrictions on the eigenvalues $\lambda_j(z)$.
We simply consider the action of the operator
$B^g(z)$  \eqref{Bg-ND} on an arbitrary Bethe vector $\mathbb{B}_{a,b}(\bu;\bv)$ using \eqref{hB-B} and action formulas \eqref{act13}--\eqref{act23}.
We also replace the expression for $B^g(z)$ \eqref{Bg-ND} by $T_{23}(z_2)\widehat{T}_{13}(z_1)-T_{13}(z_2)\widehat{T}_{12}(z_1)$ and
consider the limit $z_k\to z$ ($k=1,2$) in the end of the calculations. Then we specify  the obtained result to the semi-on-shell Bethe vectors
described in section~\ref{S-MR}.

\subsection{Action of $T_{23}(z_2)\widehat{T}_{13}(z_1)$}

In this section we study the action of
$T_{23}(z_2)\widehat{T}_{13}(z_1)$:
\be{L1}
\Lambda_1=T_{23}(z_2)\widehat{T}_{13}(z_1)\mathbb{B}_{a,b}(\bu;\bv).
\ee
Using \eqref{hB-B} we have
\be{L1-1}
\Lambda_1=(-1)^{a+b+ab}\frac{\lambda_1(\bu)\lambda_3(\bv)}{\lambda_2(\bu)\lambda_2(\bv)}T_{23}(z_2)\widehat{T}_{13}(z_1)\widehat{\mathbb{B}}_{b,a}(\bv+c;\bu).
\ee
Then due to \eqref{act13} we obtain
\be{L1-2}
\Lambda_1=\frac{(-1)^{a+b+ab}\lambda_1(\bu)\lambda_3(\bv)\hat\lambda_2(z_1)}{h(z_1,\bv+c)h(\bu,z_1)\lambda_2(\bu)\lambda_2(\bv)}
T_{23}(z_2)\widehat{\mathbb{B}}_{b+1,a+1}(\{\bv+c,z_1\};\{\bu,z_1\}).
\ee
Turning back from $\widehat{\mathbb{B}}$ to $\mathbb{B}$ and using $\hat\lambda_2(z)=\lambda_1(z)\lambda_3(z-c)$
we arrive at
\be{L1-4}
\Lambda_1=\frac{(-1)^{a+b+1}\lambda_2(z_1)\lambda_2(z_1-c)g(z_1,\bv)}{h(\bu,z_1)}
T_{23}(z_2)\mathbb{B}_{a+1,b+1}(\{\bu,z_1\};\{\bv,z_1-c\}).
\ee

It remains to act with $T_{23}(z_2)$ onto the obtained vector via \eqref{act23}:
\begin{multline}\label{L1-5}
\Lambda_1=\frac{(-1)^{a+b+1}\lambda_2(z_1)\lambda_2(z_2)\lambda_2(z_1-c)g(z_1,\bv)g(z_1,z_2)}{h(\bu,z_1)h(z_2,\bu)h(z_2,z_1)h(\bv,z_2)}\\
\times\sum_{\#\bet_{\so}=1} \frac{f(\bet_{\so},\bet_{\st})h(\bv,\bet_{\so})h(z_2,\bet_{\so})}{h(\bet_{\so},z_2)g(z_1,\bet_{\so})}
\mathbb{B}_{a+1,b+2}(\bet_{\st};\{\bv,z_1-c,z_2\}),
\end{multline}
where $\bet=\{\bu,z_1,z_2\}$ and the sum is taken over partitions $\bet\Rightarrow\{\bet_{\so},\bet_{\st}\}$ so that $\#\bet_{\so}=1$.

We see that $\bet_{\so}\ne z_1$ due to the function $g(z_1,\bet_{\so})$ in the denominator of \eqref{L1-5}. Thus, either
$\bet_{\so}=z_2$ or $\bet_{\so}=u_j$, where $j=1,\dots,a$. Respectively, we can present $\Lambda_1$ in the following form
\be{L1-pres}
\Lambda_1=\Lambda_1^{(0)}+\sum_{j=1}^a \Lambda_1^{(j)}.
\ee
Here $\Lambda_1^{(0)}$ corresponds to the case $\bet_{\so}=z_2$:
\begin{equation}\label{L10}
\Lambda_1^{(0)}=\frac{\lambda_2(z_1)\lambda_2(z_2)\lambda_2(z_1-c)g(\bv,z_1)g(\bu,z_2)g(z_1,z_2)}{h(\bu,z_1)}
\;\mathbb{B}_{a+1,b+2}(\{\bu,z_1\};\{\bv,z_1-c,z_2\}).
\end{equation}
The contributions $\Lambda_1^{(j)}$ correspond to the case $\bet_{\so}=u_j$ and have the form:
\begin{multline}\label{L1i}
\Lambda_1^{(j)}=\frac{(-1)^{a+b+1}\lambda_2(z_1)\lambda_2(z_2)\lambda_2(z_1-c)g(z_1,\bv)g(z_1,z_2)}{h(\bu,z_1)h(z_2,\bu)h(z_2,z_1)h(\bv,z_2)}\\
\times f(u_j,\bu_j)h(u_j,z_1)h(\bv,u_j)f(z_2,u_j)
\mathbb{B}_{a+1,b+2}(\{\bu_j,z_1,z_2\};\{\bv,z_1-c,z_2\}).
\end{multline}

Due to \eqref{act13} we can present the vector $\mathbb{B}_{a+1,b+2}(\{\bu_j,z_1,z_2\};\{\bv,z_1-c,z_2\})$ as a result of the  $T_{13}(z_2)$ action:
\begin{multline}\label{B-actT13}
\mathbb{B}_{a+1,b+2}(\{\bu_j,z_1,z_2\};\{\bv,z_1-c,z_2\})=\frac{h(z_2,\bu_j)h(z_2,z_1)h(\bv,z_2)}{\lambda_2(z_2)g(z_1,z_2)}\\
\times T_{13}(z_2)\mathbb{B}_{a,b+1}(\{\bu_j,z_1\};\{\bv,z_1-c\}).
\end{multline}
Then
\begin{multline}\label{L1i-1}
\Lambda_1^{(j)}=\frac{(-1)^{a+b+1}\lambda_2(z_1)\lambda_2(z_1-c)g(z_1,\bv)}{h(\bu,z_1)}h(u_j,z_1)g(z_2,u_j)f(u_j,\bu_j)h(\bv,u_j)\\
\times
 T_{13}(z_2)\mathbb{B}_{a,b+1}(\{\bu_j,z_1\};\{\bv,z_1-c\}).
\end{multline}
Observe that here we can take the limit $z_1=z_2=z$:
\begin{multline}\label{L1i-2}
\Lambda_1^{(j)}\Bigr|_{z_1=z_2=z}=\frac{(-1)^{a}\lambda_2(z)\lambda_2(z-c)g(\bv,z)}{h(\bu,z)}f(u_j,z)f(u_j,\bu_j)h(\bv,u_j)\\
\times
 T_{13}(z)\mathbb{B}_{a,b+1}(\{\bu_j,z\};\{\bv,z-c\}).
\end{multline}

\subsection{Action of $T_{13}(z_2)\widehat{T}_{12}(z_1)$}

Now we study the action of
$T_{13}(z_2)\widehat{T}_{12}(z_1)$:
\be{L2}
\Lambda_2=T_{13}(z_2)\widehat{T}_{12}(z_1)\mathbb{B}_{a,b}(\bu;\bv).
\ee
Using again \eqref{hB-B} we have
\be{L2-1}
\Lambda_2=(-1)^{a+b+ab}\frac{\lambda_1(\bu)\lambda_3(\bv)}{\lambda_2(\bu)\lambda_2(\bv)}T_{13}(z_2)\widehat{T}_{12}(z_1)\widehat{\mathbb{B}}_{b,a}(\bv+c;\bu),
\ee
and due to \eqref{act12} we obtain
\begin{multline}\label{L2-2}
\Lambda_2=\frac{(-1)^{a+b+ab}\hat\lambda_2(z_1)\lambda_1(\bu)\lambda_3(\bv)g(z_1,\bv)}
{\lambda_2(\bu)\lambda_2(\bv)h(\bu,z_1)}T_{13}(z_2)\\
\times \sum_{\#\bxi_{\so}=1}
\frac{f(\bxi_{\st},\bxi_{\so})h(\bxi_{\so},z_1)}{g(\bxi_{\so},\bv)h(z_1,\bxi_{\so})}\widehat{\mathbb{B}}_{b+1,a}(\{\bv+c,z_1\};\bxi_{\st}),
\end{multline}
where  $\bxi=\{\bu,z_1\}$ and  the sum is taken over partitions $\bxi\Rightarrow\{\bxi_{\so},\bxi_{\st}\}$ so that $\#\bxi_{\so}=1$.
Turning back to the vector $\mathbb{B}$ we find
\begin{multline}\label{L2-3}
\Lambda_2=\frac{(-1)^{a+1}\lambda_2(z_1)\lambda_2(z_1-c)g(z_1,\bv)}{h(\bu,z_1)}T_{13}(z_2)\\
\times \sum_{\#\bxi_{\so}=1} \frac{\lambda_1(\bxi_{\so}) f(\bxi_{\st},\bxi_{\so})h(\bxi_{\so},z_1)}{\lambda_2(\bxi_{\so})g(\bxi_{\so},\bv)h(z_1,\bxi_{\so})}
\mathbb{B}_{a,b+1}(\bxi_{\st};\{\bv,z_1-c\}).
\end{multline}

There is no problem to compute the action of $T_{13}(z_2)$, however, we do not do this. Instead we present the obtained result in the form similar to
\eqref{L1-pres}
\be{L2-pres}
\Lambda_2=\Lambda_2^{(0)}+\sum_{j=1}^a \Lambda_2^{(j)}.
\ee
Here $\Lambda_2^{(0)}$ corresponds to the partition $\bxi_{\so}=z_1$:
\begin{equation}\label{L20}
\Lambda_2^{(0)}=(-1)^{a+1}\lambda_1(z_1)\lambda_2(z_1-c)g(\bu,z_1)T_{13}(z_2)
\mathbb{B}_{a,b+1}(\bu;\{\bv,z_1-c\}).
\end{equation}
Observe that here we can take the limit $z_1=z_2=z$:
\begin{equation}\label{L20-1}
\Lambda_2^{(0)}\Bigr|_{z_1=z_2=z}=(-1)^{a+1}\lambda_1(z)\lambda_2(z-c)g(\bu,z)T_{13}(z)
\mathbb{B}_{a,b+1}(\bu;\{\bv,z-c\}).
\end{equation}

The contributions $\Lambda_2^{(j)}$ correspond to the partitions $\bxi_{\so}=u_j$ and have the following form:
\begin{multline}\label{L2i}
\Lambda_2^{(j)}=\frac{(-1)^{a}\lambda_2(z_1)\lambda_2(z_1-c)g(z_1,\bv)}{h(\bu,z_1)}T_{13}(z_2)\\
\times  \frac{\lambda_1(u_j)f(\bu_j,u_j) f(u_j,z_1)}{\lambda_2(u_j)g(u_j,\bv)}
\mathbb{B}_{a,b+1}(\{\bu_j,z_1\};\{\bv,z_1-c\}).
\end{multline}
Here we also can take the limit
\begin{multline}\label{L2i-1}
\Lambda_2^{(j)}\Bigr|_{z_1=z_2=z}=\frac{(-1)^{a}\lambda_2(z)\lambda_2(z-c)g(z,\bv)}{h(\bu,z)}T_{13}(z)\\
\times  \frac{\lambda_1(u_j)f(\bu_j,u_j) f(u_j,z)}{\lambda_2(u_j)g(u_j,\bv)}
\mathbb{B}_{a,b+1}(\{\bu_j,z\};\{\bv,z-c\}).
\end{multline}

\subsection{Action of $B^g(z)$ on semi-on-shell Bethe vectors\label{SS-ABgSOSBV}}

Consider the difference of the contributions $\Lambda_1^{(j)}$ and $\Lambda_2^{(j)}$ at $z_1=z_2=z$. Using \eqref{L1i-2} and \eqref{L2i-1}
we find
\begin{multline}\label{L12i}
\Bigl(\Lambda_1^{(j)}-\Lambda_2^{(j)}\Bigr)\Bigr|_{z_1=z_2=z}=(-1)^{a}\lambda_2(z)\lambda_2(z-c)
\frac{f(u_j,z)g(\bv,z)}{g(\bv,u_j)h(\bu,z)}\\[5pt]
\times\Bigl\{f(u_j,\bu_j)f(\bv,u_j)-\frac{\lambda_1(u_j)}{\lambda_2(u_j)}f(\bu_j,u_j)\Bigr\} T_{13}(z)\mathbb{B}_{a,b+1}(\{\bu_j,z\};\{\bv,z-c\}).
\end{multline}
If $\mathbb{B}_{a,b}(\bu;\bv)$ is a semi-on-shell Bethe vector such that
\be{Semi-gen}
\frac{\lambda_1(u_j)}{\lambda_2(u_j)}f(\bu_j,u_j)=f(u_j,\bu_j)f(\bv,u_j),
\ee
then this difference vanishes.
In particular, if we  impose the constraint \eqref{constV} (setting $\lambda_1(u_j)=\lambda(u_j)$ and $\lambda_2(u_j)=\kappa$),
then the contributions $\Lambda_1^{(j)}$ and $\Lambda_2^{(j)}$ cancel each other. It is remarkable, however, that the cancellation
of these terms takes place in the most general case of the semi-on-shell Bethe vectors, for which $\lambda_1(z)$ and $\lambda_2(z)$
are free functional parameters.

\section{Proof of proposition~\ref{P-Bg-BV0}\label{S-PP}}

Now we are able to prove proposition~\ref{P-Bg-BV0} via induction over $a$. For this, we specify the action formulas of
section~\ref{S-ABgBV} to the case $\lambda_1(z)=\lambda(z)$ and $\lambda_2(z)=\kappa$.

\subsection{Inductive basis \label{SS-IB}}

Consider the action of $B^g(z)$ onto $|0\rangle=\mathbb{B}_{0,0}(\emptyset;\emptyset)$. Then due to
\eqref{L10}, \eqref{L20-1} we have
\begin{equation}\label{L0vac}
\begin{aligned}
&\Lambda_1^{(0)}=\kappa^3g(z_1,z_2)\;\mathbb{B}_{1,2}(z_1;\{z_1-c,z_2\})\Bigr|_{z_1=z_2=z},\\
&\Lambda_2^{(0)}=-\kappa\lambda(z)T_{13}(z)\mathbb{B}_{0,1}(\emptyset; z-c).
\end{aligned}
\end{equation}
Using \eqref{BVt-expl1} we easily find $\mathbb{B}_{0,1}(\emptyset;z-c)=\beta\kappa^{-1}|0\rangle$. In the case
$a=1$, $b=2$, equation \eqref{BVt-expl1} gives
\begin{equation}\label{BVt12}
\mathbb{B}_{1,2}(u;\{v_1,v_2\})=\frac{\beta^{2}}{\kappa^{3}g(\bv,u)}\left(
T_{12}(u)|0\rangle +\frac\kappa\beta(f(\bv,u)-1) T_{13}(u)|0\rangle \right).
\end{equation}
Setting here $u=z_1$, $v_1=z_1-c$, and $v_2=z_2$ we obtain
\begin{equation}\label{BVt12zz}
\mathbb{B}_{1,2}(z_1;\{z_1-c,z_2\})=\frac{\beta^{2}}{\kappa^{3}g(z_1,z_2)}\left(
T_{12}(z_1)|0\rangle -\frac\kappa\beta T_{13}(z_1)|0\rangle \right),
\end{equation}
leading to
\be{Lam1per}
\Lambda_1^{(0)}=\beta^{2}T_{12}(z)|0\rangle -\kappa\beta T_{13}(z)|0\rangle.
\ee
Thus, we arrive at
\be{actBg1}
B^g(z)|0\rangle=\Lambda_1^{(0)}-\Lambda_2^{(0)}=\beta^{2}T_{12}(z)|0\rangle +\beta(\lambda(z)-\kappa) T_{13}(z)|0\rangle.
\ee
It is easy to see that representation \eqref{Bg-Maction} gives the same result for $a=1$:
\begin{equation}\label{BV1}
B^g(z)|0\rangle=
\sum_{n=0}^1\beta^{2-n}\sum_{s=0}^n\sum_{\substack{\#\bar z_{\so}=s\\\#\bar z_{\st}=n-s}}
(-\kappa)^{n-s}\lambda(\bar z_{\so}) T_{13}(\bar z_{\so})T_{13}(\bar z_{\st})T_{12}(\bar z_{\sth})|0\rangle.
\end{equation}
Here the sum is taken over partitions of the set $\bar z$ (consisting on one element $z$) into three subsets
$\bar z_{\so}$, $\bar z_{\st}$, and $\bar z_{\sth}$. Clearly, two of these subsets are empty. Because of this reason
we did not write the product of the $f$-functions in \eqref{BV1} (see \eqref{Bg-Maction}), as these products are
taken at least over one empty set. Setting successively in \eqref{BV1} $\bar z_{\so}=z$, $\bar z_{\st}=z$, and $\bar z_{\sth}=z$
we obtain three contributions coinciding with \eqref{actBg1}.
Thus, the  induction basis is checked.

It is interesting to write down this result in terms of the entries of the original monodromy matrix $T^0(u)$. Using
\eqref{TwMM} and \eqref{gf} we find
\be{T0T}
T_{13}(u)=T^0_{13}(u)- \frac\beta{1-\kappa}T^0_{12}(u), \qquad T_{12}(u)=T^0_{12}(u).
\ee
Then replacing $z$ with $u$ in representation \eqref{actBg1} we obtain
\be{Bg1T0}
B^g(u)|0\rangle=\beta(\lambda(u)-\kappa) T^0_{13}(u)|0\rangle+
\beta^{2}\left(\frac{1-\lambda(u)}{1-\kappa}\right)T^0_{12}(u)|0\rangle.
\ee

The monodromy matrix $T^0(u)$ has two on-shell Bethe vectors in the case $a=1$: $B^0_{1,0}(u,\emptyset)$ and $B^0_{1,1}(u,v)$.
In the first case, there is only one Bethe
equation  $\lambda(u)=\kappa$, and hence, \eqref{Bg1T0} yields
\be{Bg1T010}
B^g(u)|0\rangle=\beta^{2}T^0_{12}(u)|0\rangle.
\ee
In the second case we have a system of two Bethe equations
\be{sysBE}
\lambda(u)=\kappa f(v,u), \qquad \kappa f(v,u)=1,
\ee
what implies $\lambda(u)=1$. Then \eqref{Bg1T0} yields
\be{Bg1T011}
B^g(u)|0\rangle=\beta(1-\kappa) T^0_{13}(u)|0\rangle.
\ee
Both vectors $ T^0_{12}(u)|0\rangle$ and $ T^0_{13}(u)|0\rangle$ indeed are on-shell Bethe vectors
respectively for $\lambda(u)=\kappa$ and $\lambda(u)=1$. Thus, the action of $B^g(u)$ onto the
pseudovacuum vector does give the on-shell Bethe vectors, if $u$ is a root of Bethe equations.

\subsection{Inductive step \label{SS-IS}}

We assume that \eqref{Bg-Maction} holds for some $a\ge 1$. Then due to corollary~\ref{P-Bg-BV} the action
$B^g(\bu)|0\rangle$ with $\bu=\{u_1,\dots,u_a\}$ is proportional to the semi-on-shell Bethe vector $\mathbb{B}_{a,b}(\bu,\bv)$
\eqref{actBg}, where the set $\bv$ enjoys
the constraint \eqref{constV}. Hence, we have
\be{actBgz}
B^g(z)B^g(\bu)|0\rangle=\beta^{2a-b}\kappa^{a+b} g(\bv,\bu)B^g(z)\mathbb{B}_{a,b}(\bu,\bv).
\ee
The action of $B^g(z)$ onto  $\mathbb{B}_{a,b}(\bu,\bv)$
is given by the terms
$\Lambda_2^{(0)}$ \eqref{L20-1} and $\Lambda_1^{(0)}$ \eqref{L10} (in the  limit $z_1=z_2=z$). Thus,
\be{actBgz1}
B^g(z)B^g(\bu)|0\rangle=\beta^{2a-b}\kappa^{a+b} g(\bv,\bu)\Bigl(\Lambda_1^{(0)}-\Lambda_2^{(0)}\Bigr)\Bigr|_{z_1=z_2=z}.
\ee
Now we should set $\lambda_1(z)= \lambda(z)$ and $\lambda_2(z)=\kappa$ in \eqref{L10} and \eqref{L20-1} for $\Lambda_k^{(0)}$
and substitute these expressions into \eqref{actBgz1}. We obtain
\be{Bg-newB}
B^g(z)B^g(\bu)|0\rangle=M_1 + M_2,
\ee
where
\begin{equation}\label{M1}
M_1=\beta^{2a-b}\kappa^{a+b+1} g(\bv,\bu)
\lambda(z) g(z,\bu)T_{13}(z)\mathbb{B}_{a,b+1}(\bu;\{\bv,z-c\}),
\end{equation}
and
\begin{equation}\label{M2}
M_2=\beta^{2a-b}\kappa^{a+b+3} g(\bv,\bu)
\frac{g(\bv,z_1)g(\bu,z_2)g(z_1,z_2)}{h(\bu,z_1)}
\;\mathbb{B}_{a+1,b+2}(\{\bu,z_1\};\{\bv,z_1-c,z_2\})\Bigr|_{z_1=z_2=z}.
\end{equation}
It remains to substitute explicit expression \eqref{BVt-expl1} for the Bethe vectors  $\mathbb{B}_{a,b+1}(\bu;\{\bv,z-c\})$ and
$\mathbb{B}_{a+1,b+2}(\{\bu,z_1\};\{\bv,z_1-c,z_2\})$ into \eqref{M1} and \eqref{M2}. However, before doing this, it is convenient to
describe an expected form of the result.

We expect that multiple action $B^g(z)B^g(\bu)|0\rangle$ is given by \eqref{Bg-Maction}, in which one should replace
$\bu$ with $\bet=\{\bu,z\}$ and $a\to a+1$. That is,
\begin{multline}\label{BVa1}
B^g(z)B^g(\bu)|0\rangle=
\sum_{n=0}^{a+1}\beta^{2a+2-n}\sum_{s=0}^n\sum_{\substack{\#\bet_{\so}=s\\\#\bet_{\st}=n-s}}
(-\kappa)^{n-s}\lambda(\bet_{\so})f(\bet_{\st},\bet_{\so})f(\bet_{\sth},\bet_{\so})f(\bet_{\st},\bet_{\sth})\\
\times T_{13}(\bet_{\so})T_{13}(\bet_{\st})T_{12}(\bet_{\sth})|0\rangle.
\end{multline}
Let us give more details on the expected form of the result \eqref{BVa1}.

There are three possibilities in the sum over partitions in the rhs of \eqref{BVa1}: $z\in\bet_{\so}$;
$z\in\bet_{\st}$; $z\in\bet_{\sth}$. Respectively, there are three contributions
\be{B-B3}
B^g(z)B^g(\bu)|0\rangle=W_1+W_2+W_3.
\ee
In the first case we have
\begin{multline}\label{B1}
W_1=\lambda(z)T_{13}(z)\sum_{n=0}^{a}\beta^{2a+1-n}\sum_{s=0}^n\sum_{\substack{\#\bu_{\so}=s\\\#\bu_{\st}=n-s}}
(-\kappa)^{n-s}\lambda(\bu_{\so})f(\bu_{\st},\bu_{\so})f(\bu_{\sth},\bu_{\so})f(\bu_{\st},\bu_{\sth})\\
\times f(\bu_{\st},z)f(\bu_{\sth},z) T_{13}(\bu_{\so})T_{13}(\bu_{\st})T_{12}(\bu_{\sth})|0\rangle.
\end{multline}
Indeed, we can set $\bet_{\so}=\{z,\bu_{\so}\}$, $\bet_{\st}=\bu_{\st}$, and $\bet_{\sth}=\bu_{\sth}$. The set
$\bet_{\so}$ is not empty, thus, $s\in[1,\dots,n]$. This implies $n\in[1,\dots,a+1]$. Shifting $n\to n+1$ and $s\to s+1$ in \eqref{BVa1}
we arrive at \eqref{B1}.

In the second case $\bet_{\st}=\{z,\bu_{\st}\}$, $\bet_{\so}=\bu_{\so}$, and $\bet_{\sth}=\bu_{\sth}$. The set $\bet_{\st}$ is not empty,
thus, $s\in[0,\dots,n-1]$. We also have $n\in[1,\dots,a+1]$, because the union $\{\bet_{\so},\bet_{\st}\}$ is not empty. Shifting
$n\to n+1$ in \eqref{BVa1} we arrive at
\begin{multline}\label{B2}
W_2=-\kappa T_{13}(z)\sum_{n=0}^{a}\beta^{2a+1-n}\sum_{s=0}^n\sum_{\substack{\#\bu_{\so}=s\\\#\bu_{\st}=n-s}}
(-\kappa)^{n-s}\lambda(\bu_{\so})f(\bu_{\st},\bu_{\so})f(\bu_{\sth},\bu_{\so})f(\bu_{\st},\bu_{\sth})\\
\times f(z,\bu_{\so})f(z,\bu_{\sth})  T_{13}(\bu_{\so})T_{13}(\bu_{\st})T_{12}(\bu_{\sth})|0\rangle.
\end{multline}

Finally, in the third case $\bet_{\so}=\bu_{\so}$, $\bet_{\st}=\bu_{\st}$, and $\bet_{\sth}=\{z,\bu_{\sth}\}$. The set
$\bet_{\sth}$ is not empty, thus, $n\in[0,\dots,a]$. We obtain
\begin{multline}\label{B3}
W_3=\sum_{n=0}^{a}\beta^{2a+2-n}\sum_{s=0}^n\sum_{\substack{\#\bu_{\so}=s\\\#\bu_{\st}=n-s}}
(-\kappa)^{n-s}\lambda(\bu_{\so})f(\bu_{\st},\bu_{\so})f(\bu_{\sth},\bu_{\so})f(\bu_{\st},\bu_{\sth})\\
\times f(z,\bu_{\so})f(\bu_{\st},z) T_{13}(\bu_{\so})T_{13}(\bu_{\st})T_{12}(\bu_{\sth})T_{12}(z)|0\rangle.
\end{multline}

Thus, our goal is to check that equations \eqref{Bg-newB}--\eqref{M2} give all three contributions $W_j$, $j=1,2,3$.

\subsubsection{Contribution $W_1$}

Consider the term $M_1$. Using \eqref{BVt-expl1} for $\mathbb{B}_{a,b+1}(\bu;\{\bv,z-c\})$ we obtain
\begin{multline}\label{B1-pre1}
M_1=\lambda(z)f(\bu,z)T_{13}(z)
\sum_{n=0}^a\kappa^n\beta^{2a+1-n}\sum_{s=0}^n\sum_{\substack{\#\bu_{\so}=s\\\#\bu_{\st}=n-s}}
(-1)^{n-s}\\
\times f(\bu_{\so},\bu_{\st})\frac{f(\bv,\bu_{\so})}{f(\bu_{\so},z)}f(\bu_{\so},\bu_{\sth})f(\bu_{\st},\bu_{\sth})
\; T_{13}(\bu_{\so})T_{13}(\bu_{\st})T_{12}(\bu_{\sth})|0\rangle.
\end{multline}
Taking into account \eqref{constV} we arrive at
\begin{multline}\label{B1-pre2}
M_1=\lambda(z)T_{13}(z)
\sum_{n=0}^a\beta^{2a+1-n}\sum_{s=0}^n\sum_{\substack{\#\bu_{\so}=s\\\#\bu_{\st}=n-s}}
(-\kappa)^{n-s}f(\bu_{\st},z)f(\bu_{\sth},z)\\
\times \lambda(\bu_{\so}) f(\bu_{\st},\bu_{\so})f(\bu_{\sth},\bu_{\so})f(\bu_{\st},\bu_{\sth})
\; T_{13}(\bu_{\so})T_{13}(\bu_{\st})T_{12}(\bu_{\sth})|0\rangle.
\end{multline}
We see that this is exactly $W_1$ \eqref{B1}.

\subsubsection{Contributions $W_2$ and $W_3$}
Consider the term $M_2$. Using \eqref{BVt-expl1} for $\mathbb{B}_{a+1,b+2}(\{\bu,z_1\};\{\bv,z_1-c,z_2\})$ and setting
$\{z_1,\bu\}=\bet$ we obtain
\begin{multline}\label{B23-pre1}
M_2=\sum_{n=0}^{a+1}\kappa^n\beta^{2a+2-n}\sum_{s=0}^n\sum_{\substack{\#\bet_{\so}=s\\\#\bet_{\st}=n-s}}
(-1)^{n-s}\frac{f(\bv,\bet_{\so})f(z_2,\bet_{\so})}{f(\bet_{\so},z_1)}\\
\times f(\bet_{\so},\bet_{\st})f(\bet_{\so},\bet_{\sth})f(\bet_{\st},\bet_{\sth})
\; T_{13}(\bet_{\so})T_{13}(\bet_{\st})T_{12}(\bet_{\sth})|0\rangle.
\end{multline}
We see that $z_1\notin\bet_{\so}$, otherwise $1/f(\bet_{\so},z_1)=0$. Thus, either $z_1\in\bet_{\st}$ or $z_1\in\bet_{\sth}$. Respectively,
$M_2$ consists of two contributions: $M_2^{(1)}$ corresponding to the case $z_1\in\bet_{\st}$ and
$M_2^{(2)}$ corresponding to the case $z_1\in\bet_{\sth}$.

Let $z_1\in\bet_{\st}$. Then we can set
$\bet_{\so}=\bu_{\so}$, $\bet_{\sth}=\bu_{\sth}$, and $\bet_{\st}=\{z_1,\bu_{\st}\}$. We also have $n-s>0$, and thus, $s\in [0,\dots,n-1]$ and
$n\in [1,\dots,a+1]$. Shifting $n\to n+1$   and setting $z_1=z_2=z$ we obtain
\begin{multline}\label{B2-pre1}
M_2^{(1)}=-\kappa T_{13}(z)
\sum_{n=0}^{a}\kappa^n\beta^{2a+1-n}\sum_{s=0}^{n}\sum_{\substack{\#\bu_{\so}=s\\\#\bu_{\st}=n-s}}
(-1)^{n-s}f(z,\bu_{\so})f(z,\bu_{\sth})\\
\times f(\bu_{\so},\bu_{\st})f(\bv,\bu_{\so})f(\bu_{\so},\bu_{\sth})f(\bu_{\st},\bu_{\sth})
\; T_{13}(\bu_{\so})T_{13}(\bu_{\st})T_{12}(\bu_{\sth})|0\rangle.
\end{multline}
Using \eqref{constV} we arrive at
\begin{multline}\label{B2-pre2}
M_2^{(1)}=-\kappa T_{13}(z)
\sum_{n=0}^{a}\beta^{2a+1-n}\sum_{s=0}^{n}\sum_{\substack{\#\bu_{\so}=s\\\#\bu_{\st}=n-s}}
(-\kappa)^{n-s}f(z,\bu_{\so})f(z,\bu_{\sth})\\
\times \lambda(\bu_{\so}) f(\bu_{\st},\bu_{\so})f(\bu_{\sth},\bu_{\so})f(\bu_{\st},\bu_{\sth})
\; T_{13}(\bu_{\so})T_{13}(\bu_{\st})T_{12}(\bu_{\sth})|0\rangle,
\end{multline}
and we see that this is exactly $W_2$ \eqref{B2}.

Let now $z_1\in\bet_{\sth}$. Then we can set
$\bet_{\so}=\bu_{\so}$, $\bet_{\st}=\bu_{\st}$, and $\bet_{\sth}=\{z_1,\bu_{\sth}\}$. We also have $n\in [0,\dots,a]$.
 Setting $z_1=z_2=z$ we obtain
\begin{multline}\label{B3-pre1}
M_2^{(2)}=
\sum_{n=0}^{a}\kappa^n\beta^{2a+2-n}\sum_{s=0}^n\sum_{\substack{\#\bu_{\so}=s\\\#\bu_{\st}=n-s}}
(-1)^{n-s}f(z,\bu_{\so})f(\bu_{\st},z)\\
\times f(\bu_{\so},\bu_{\st})f(\bv,\bu_{\so})f(\bu_{\so},\bu_{\sth})f(\bu_{\st},\bu_{\sth})
\; T_{13}(\bu_{\so})T_{13}(\bu_{\st})T_{12}(\bu_{\sth})T_{12}(z)|0\rangle.
\end{multline}
Using \eqref{constV} we arrive at
\begin{multline}\label{B3-pre2}
M_2^{(2)}=
\sum_{n=0}^{a}\beta^{2a+2-n}\sum_{s=0}^n\sum_{\substack{\#\bu_{\so}=s\\\#\bu_{\st}=n-s}}
(-\kappa)^{n-s}f(z,\bu_{\so})f(\bu_{\st},z)\\
\times \lambda(\bu_{\so}) f(\bu_{\st},\bu_{\so})f(\bu_{\sth},\bu_{\so})f(\bu_{\st},\bu_{\sth})
\; T_{13}(\bu_{\so})T_{13}(\bu_{\st})T_{12}(\bu_{\sth})T_{12}(z)|0\rangle.
\end{multline}
We see that this is exactly $W_3$ \eqref{B3}. Thus, the multiple action $B^g(z)B^g(\bu)|0\rangle$ is given by the
formulas \eqref{B1}--\eqref{B3}, leading to \eqref{BVa1}. Hence,
proposition~\ref{P-Bg-BV0} is proved.

\section*{Conclusion}

In this paper we have proved one of conjectures of \cite{GroLMS17}. Namely we have shown that the successive action
of the operator $B^g$ \eqref{Bg} on the pseudovacuum vector
generates on-shell Bethe vectors in $\mathfrak{gl}_3$-invariant models, provided the arguments of these operators
satisfy Bethe equations. Furthermore, if the arguments of the $B^g$ operators are generic complex numbers, then the
successive action of $B^g$ gives a semi-on-shell Bethe vector. This property holds not only for $\mathfrak{gl}_3$-invariant
spin chains, but for a wider class of models, for instance, for the two-component generalization of the Lieb--Liniger
model \cite{LiebL63,Lieb63,Yang67,Sla15}.
At the same time, we would like to emphasize that the operator $B^g$ can not be used to construct on-shell Bethe vectors in generic
NABA-solvable models. The restriction $T^0_{23}(u)|0\rangle=0$ is crucial.
On the other hand, the existence of this restriction was clear from the outset, since within the framework of the new approach
Bethe vectors depend  only on one set of variables by construction, rather than two sets, as is the case of the Bethe vectors of  the general form.

In this paper, we considered the minimal twist \eqref{gf}. A general twist $K^{gen}$ can be treated as further
twisting of the matrix $T(u)$. It is quite natural to expect  that the effect of the general twist must be similar to what one has in the
case of $\mathfrak{gl}_2$ based models  \cite{BelS18}. Namely, we saw that for the minimal twist, the multiple action of $B^g$ was equivalent to the {\it one}
semi-on-shell Bethe vector $\mathbb{B}_{a,b}(\bu,\bv)$. Most probably, that the multiple action of $B^g$ in the case of the general twist
is equivalent to a linear combination
of semi-on-shell Bethe vectors with different sets of the Bethe parameters. However, as soon as we impose Bethe equations, only one term in this linear
combination should survive. The proof of this property in the case of $\mathfrak{gl}_2$-invariant models is very simple (see
\cite{BelS18}). However, a generalization of this proof to  the models with $\mathfrak{gl}_3$-invariant $R$-matrix meats certain
technical difficulties. Therefore, we did not consider the case of the general twist.

Despite the fact that we have proved the property of $B^g(u)$  to generate on-shell Bethe vectors,
we still do not have a clear understanding of why this is happening. In this context,
the most intrigues looks the cancellation of `unwanted' terms \eqref{L12i}. Recall that this cancellation takes place
for  a {\it general} semi-on-shell Bethe vector. We do not need to assume any specific form
of $\lambda_j(u)$ and specific action of $T_{23}(u)$ onto $|0\rangle$. Perhaps this is due to some hidden structure of the
operator $B^g(u)$, which is not yet clear. It would be very interesting to find this structure.

Finally it is worth mentioning that a generalization of the operator $B^g(u)$ to
the $\mathfrak{gl}_N$-invariant spin chains ($N>3$) was also proposed in  \cite{GroLMS17}. It was conjectured
that this operator also generates on-shell Bethe vectors, similarly to the $\mathfrak{gl}_3$ case. Basing on the results of
this paper we can assume that the successive action of $B^g(u)$ is equivalent to a semi-on-shell Bethe vector of a certain
$\mathfrak{gl}_N$-invariant integrable model. However, the method that we used in this paper hardly can be applied  to
the case $N>3$, as it becomes very bulky.

\section*{Acknowledgements}
We would like to thank I. Kostov, F. Levkovich-Maslyuk, S. Pakuliak, and E. Ragoucy for numerous and
fruitful discussions. The work of A.L. has been funded by Russian Academic Excellence Project 5-100 and 
by Young Russian Mathematics award.

A part of this work, section~\ref{S-ABgBV}, was performed in Steklov Mathematical Institute
of Russian Academy of Sciences by N.A.S. and he was supported by the Russian Science Foundation
under grant 14-50-00005.

\appendix

\section{Properties of DWPF\label{A-DWPF}}

The DWPF $K_n(\bx|\by)$ defined by \eqref{K-def} is a rational function of $\bx$ and $\by$. It is symmetric over $\bx$ and
symmetric over $\by$. If $x_j\to\infty$ (or $y_j\to\infty$) and all other variables are fixed, then $K_n(\bx|\by)\to 0$.
This function has simple poles at $x_j=y_k$, $j,k=1,\dots,n$. The residues in these poles can be expressed in terms of $K_{n-1}$.
Due to the symmetry of $K_n$ over $\bx$ and over $\by$, it is enough to consider the residue at $x_n=y_n$:
\be{resK}
K_n(\bx|\by)\Bigr|_{x_n\to y_n}= g(x_n,y_n)f(\bx_n,x_n)f(y_n,\by_n) K_{n-1}(\bx_n|\by_n)+reg,
\ee
where $reg$ means regular part.

The properties listed above, together with the initial condition $K_1(x|y)=g(x,y)$ fix the function $K_n(\bx|\by)$ unambiguously
\cite{Kor82,Ize87}.

\begin{prop}\label{P-A1}
Let $\#\bx=m$ and $\#\by=n$ so that $m\ge n$. Then
\be{ident}
\sum_{\#\bx_{\so}=n} K_n(\bx_{\so}|\by)f(\bx_{\st},\bx_{\so})=\sum_{k=0}^n\sum_{\#\by_{\so}=k} (-1)^{n-k}f(\by_{\so},\by_{\st})f(\bx,\by_{\so}).
\ee
Here in the lhs, the sum is taken over partitions $\bx\Rightarrow\{\bx_{\so},\bx_{\st}\}$ so that $\#\bx_{\so}=n$.
In the rhs, the sum is taken over all possible partitions $\by\Rightarrow\{\by_{\so},\by_{\st}\}$.
\end{prop}

{\sl Proof.} We use induction over $n$. For $n=1$, equation \eqref{ident} takes the form
\be{identn1}
\sum_{j=1}^m g(x_j,y)f(\bx_j,x_j)=f(\bx,y)-1.
\ee
Obviously, the lhs of \eqref{identn1} is partial fraction decomposition of the rhs. Thus, identity \eqref{ident} is valid
for $n=1$ and arbitrary $m\ge 1$.

Assume that \eqref{ident} holds for some $n-1$ and arbitrary $m\ge n-1$. Let
\be{Hlr}
\begin{aligned}
&H_{n,m}^{\ell}(\bx;\by)= \sum_{\#\bx_{\so}=n} K_n(\bx_{\so}|\by)f(\bx_{\st},\bx_{\so}),\\
&H_{n,m}^{r}(\bx;\by)= \sum_{k=0}^n\sum_{\#\by_{\so}=k} (-1)^{n-k}f(\by_{\so},\by_{\st})f(\bx,\by_{\so}).
\end{aligned}
\ee
Consider properties of $H_{n,m}^{\ell}$ and $H_{n,m}^{r}$ as functions of $y_n$ at other variables fixed.
Both functions are rational functions of $y_n$. Due to the properties of $K_n(\bx_{\so}|\by)$, the function
$H_{n,m}^{\ell}(\bx;\by)$ vanishes as $y_n\to\infty$. Let us show that $H_{n,m}^{r}(\bx;\by)$ has the same
property. We use the fact that for arbitrary finite $z$ the functions
$f(z,y_n)$ and $f(y_n,z)$ go to $1$ as $y_n\to\infty$.

Clearly, we have either $y_n\in\by_{\so}$ or $y_n\in\by_{\st}$ in the sum over partitions over $\by$. Consider the first case.
Then $k>0$ and we can set $\by_{\so}=\{y_n,\by_{\qo}\}$. We obtain
\be{1stcon}
\lim_{y_n\to\infty}\sum_{k=1}^n\sum_{\#\by_{\qo}=k-1}\!\! (-1)^{n-k}f(y_n,\by_{\st}) f(\by_{\qo},\by_{\st})f(\bx,y_n)f(\bx,\by_{\qo})=
\sum_{k=0}^{n-1}\sum_{\#\by_{\qo}=k} (-1)^{n-k-1}f(\by_{\qo},\by_{\st})f(\bx,\by_{\qo}).
\ee
In the second case $k<n$ and we can set $\by_{\st}=\{y_n,\by_{\qt}\}$. We obtain
\be{2ndcon}
\lim_{y_n\to\infty}\sum_{k=0}^{n-1}\sum_{\#\by_{\so}=k} (-1)^{n-k}f(\by_{\so},y_n)f(\by_{\so},\by_{\qt})f(\bx,\by_{\so})
=\sum_{k=0}^{n-1}\sum_{\#\by_{\so}=k} (-1)^{n-k}f(\by_{\so},\by_{\qt})f(\bx,\by_{\so}).
\ee
Relabeling $\by_{\qo}\to \by_{\so}$ in \eqref{1stcon} and $\by_{\qt}\to \by_{\st}$ in \eqref{2ndcon} we see that the obtained sums
over partitions cancel each other. Thus $H_{n,m}^{r}(\bx;\by)\to 0$ as $y_n\to\infty$.

It remains to compare the residues of two rational functions in the poles $y_n=x_j$, $j=1,\dots,m$. Let $y_n\to x_j$ in the function
$H_{n,m}^{\ell}(\bx;\by)$. The pole occurs if and only if $x_j\in\bx_{\so}$. Setting $\bx_{\so}=\{x_j,\bx_{\so'}\}$ and using
\eqref{resK} we find
\be{Hlpole1}
H_{n,m}^{\ell}(\bx;\by)\Bigr|_{y_n\to x_j}= \sum_{\#\bx_{\so'}=n-1} g(x_j,y_n)f(\bx_{\so'},x_j)f(y_n,\by_n)K_{n-1}(\bx_{\so'}|\by_n)
f(\bx_{\st},\bx_{\so'})f(\bx_{\st},x_j) + reg,
\ee
where $reg$ means regular part. Obviously $f(\bx_{\so'},x_j)f(\bx_{\st},x_j)=f(\bx_j,x_j)$. Hence,
\be{Hlpole2}
H_{n,m}^{\ell}(\bx;\by)\Bigr|_{y_n\to x_j}= g(x_j,y_n)f(\bx_j,x_j)f(y_n,\by_n)\sum_{\#\bx_{\so'}=n-1} K_{n-1}(\bx_{\so'}|\by_n)
f(\bx_{\st},\bx_{\so'}) + reg.
\ee
The remaining sum over partitions gives $H_{n-1,m-1}^{\ell}(\bx_j;\by_n)$, and we finally arrive at
\be{Hlpole3}
H_{n,m}^{\ell}(\bx;\by)\Bigr|_{y_n\to x_j}= g(x_j,y_n)f(\bx_j,x_j)f(y_n,\by_n)H_{n-1,m-1}^{\ell}(\bx_j;\by_n)+ reg.
\ee

Consider now the behavior of $H_{n,m}^{r}(\bx;\by)$ at $y_n\to x_j$. The pole occurs if and only if $y_n\in\by_{\so}$. Setting $\by_{\so}=\{y_n,\by_{\so'}\}$
we obtain
\begin{multline}\label{Hrpole1}
H_{n,m}^{r}(\bx;\by)\Bigr|_{y_n\to x_j}= \sum_{k=1}^n\sum_{\#\by_{\so'}=k-1} (-1)^{n-k}f(y_n,\by_{\st})f(\by_{\so'},\by_{\st})\\
\times f(\bx_j,\by_{\so'}) g(x_j,y_n)f(\bx_j,x_j)f(y_n,\by_{\so'})+reg.
\end{multline}
Using $f(y_n,\by_{\so'})f(y_n,\by_{\st})=f(y_n,\by_n)$ and changing $k\to k+1$ we find
\begin{equation}\label{Hrpole2}
H_{n,m}^{r}(\bx;\by)\Bigr|_{y_n\to x_j}= g(x_j,y_n)f(\bx_j,x_j)f(y_n,\by_n)\sum_{k=0}^{n-1}\sum_{\#\by_{\so'}=k} (-1)^{n-1-k}
f(\by_{\so'},\by_{\st})f(\bx_j,\by_{\so'}) +reg.
\end{equation}
The remaining sum over partitions gives $H_{n-1,m-1}^{r}(\bx_j;\by_n)$, and we finally arrive at
\begin{equation}\label{Hrpole3}
H_{n,m}^{r}(\bx;\by)\Bigr|_{y_n\to x_j}= g(x_j,y_n)f(\bx_j,x_j)f(y_n,\by_n)H_{n-1,m-1}^{r}(\bx_j;\by_n) +reg.
\end{equation}
Due to the induction assumption $H_{n-1,m-1}^{r}(\bx_j;\by_n)=H_{n-1,m-1}^{\ell}(\bx_j;\by_n)$. Hence, the residues of
$H_{n,m}^{r}(\bx;\by)$ and $H_{n,m}^{\ell}(\bx;\by)$ in the poles at $y_n=x_j$ coincide. Since both functions vanish at
$y_n\to\infty$ we conclude that $H_{n,m}^{r}(\bx;\by)=H_{n,m}^{\ell}(\bx;\by)$. \qed

\begin{prop}\label{P-A12}
Let $\#\bx=m$ and $\#\by=n$. Then
\be{identA2}
\sum_{k=0}^n\sum_{\#\by_{\so}=k} (-1)^{n-k}f(\by_{\so},\by_{\st})f(\bx,\by_{\so})
=\det_n\left(\frac{f(y_j,\by_j)f(\bx,y_j)}{h(y_j,y_k)}-\delta_{jk}\right).
\ee
Here the sum is taken over all possible partitions $\by\Rightarrow\{\by_{\so},\by_{\st}\}$.
If $m<n$, then
\be{identA3}
\det_n\left(\frac{f(y_j,\by_j)f(\bx,y_j)}{h(y_j,y_k)}-\delta_{jk}\right)=0.
\ee
\end{prop}

{\sl Proof}. Expanding the determinant in the rhs of  \eqref{identA2} over diagonal minors we find
\begin{multline}\label{Exp-det}
\det_n\left(\frac{f(y_j,\by_j)f(\bx,y_j)}{h(y_j,y_k)}-\delta_{jk}\right)\\
=(-1)^{n}
+\sum_{s=1}^{n}(-1)^{n-s}\sum_{1\le j_1<\dots<j_s\le n}\left(\prod_{p=1}^s f(y_{j_p},\by_{j_p})f(\bx,y_{j_p})\right)\det_{s}\frac1{h(y_{j_i},y_{j_k})}.
\end{multline}
The determinant in the rhs is the Cauchy determinant, hence,
\be{Cau}
\det_{s}\frac1{h(y_{j_i},y_{j_k})}=\prod_{\substack{p,q=1\\ p\ne q}}^s\frac1{f(y_{j_p},y_{j_q})}.
\ee
Thus, we obtain
\begin{multline}\label{Exp-det1}
\det_n\left(\frac{f(y_j,\by_j)f(\bx,y_j)}{h(y_j,y_k)}-\delta_{jk}\right)\\
=(-1)^{n}+\sum_{s=1}^{n}(-1)^{n-s}\sum_{1\le j_1<\dots<j_s\le n}\left(\prod_{p=1}^s
f(y_{j_p},\by_{j_p})f(\bx,y_{j_p})\right)
\prod_{\substack{p,q=1\\ p\ne q}}^s\frac1{f(y_{j_p},y_{j_q})}.
\end{multline}
This is exactly the sum over partitions in the lhs of \eqref{identA2}.

Let now $m<n$. Obviously,
\be{Simtr}
\det_n\left(\frac{f(y_j,\by_j)f(\bx,y_j)}{h(y_j,y_k)}-\delta_{jk}\right)=
\det_n\left(\frac{f(y_j,\by_j)f(\bx,y_k)}{h(y_j,y_k)}-\delta_{jk}\right),
\ee
because both matrices are related by a similarity transformation. It is easy to see that
the matrix in the rhs of \eqref{Simtr} has an eigenvector with zero eigenvalue:
\be{zeroEV}
\sum_{k=1}^n \frac{f(y_j,\by_j)f(\bx,y_k)}{h(y_j,y_k)}\,\nu_k-\nu_j=0,
\ee
where
\be{nu}
\nu_k=\frac{g(y_k,\by_k)}{g(\bx,y_k)}.
\ee
Indeed, consider a function
\be{funct}
\frac{1}{h(z,\by)g(\bx,z)}=\frac{c^{n-m}\prod_{p=1}^m(x_p-z)}{\prod_{q=1}^n(z-y_q+c)}.
\ee
Due to the condition $m<n$ this function vanishes as $z\to\infty$. Hence, it has the following partial fraction decomposition
\be{FD1}
\frac{c^{n-m}\prod_{p=1}^m(x_p-z)}{\prod_{q=1}^n(z-y_q+c)}=\sum_{k=1}^n\frac{c^{n-m}\prod_{p=1}^m(x_p-y_k+c)}{(z-y_k+c)\prod_{q=1, q\ne k}^n(y_k-y_q)}=
\sum_{k=1}^n \frac{g(y_k,\by_k)h(\bx,y_k)}{h(z,y_k)}.
\ee
Setting here $z=y_j$ we arrive at
\be{FD2}
\sum_{k=1}^n \frac{g(y_k,\by_k)h(\bx,y_k)}{h(y_j,y_k)}- \frac{1}{h(y_j,\by)g(\bx,y_j)}=0.
\ee
On the other hand, substituting $\nu_j$ from \eqref{nu} into \eqref{zeroEV} we immediately obtain the lhs of \eqref{FD2}.\qed

\section{Proof of the connection between two types of Bethe vectors\label{S-PCBTTBV}}

The proof of \eqref{hB-B} is based on the double induction, first on $a$, and then on $b$.

\subsection{First step of induction\label{S-FSI}}

We first assume that $b=0$. Then \eqref{hB-B} takes the form
\begin{equation}\label{two-BVnorm}
   \widehat{\mathbb{B}}_{0,a}(\emptyset;\bu)  = (-1)^a\frac{\lambda_2(\bu)}{\lambda_1(\bu) }   \mathbb{B}_{a,0}(\bu;\emptyset).
\end{equation}
For $a=0$, \eqref{two-BVnorm} turns into a trivial identity: $|0\rangle=|0\rangle$. It is easy to see that \eqref{two-BVnorm} also holds for $a=1$:
\begin{equation}\label{BVa11}
   \widehat{\mathbb{B}}_{0,1}(\emptyset;u)  = \frac{\widehat{T}_{23}(u)|0\rangle}{\hat\lambda_2(u) } =
-\frac{T_{12}(u)|0\rangle}{\lambda_1(u) } =-\frac{\lambda_2(u)}{\lambda_1(u) } \mathbb{B}_{1,0}(\bu;\emptyset),
\end{equation}
where we used \eqref{HT} for $\widehat{T}_{23}(u)$.

Assume now that \eqref{two-BVnorm} holds for some $a\ge 1$. Then we have for $\#\bu=a$
\begin{equation}\label{BVap10}
   \widehat{\mathbb{B}}_{0,a+1}(\emptyset;\{\bu,z\})  = \widehat{T}_{23}(z)\frac{\widehat{T}_{23}(\bu)|0\rangle}{\hat\lambda_2(z)\hat\lambda_2(\bu) } =
(-1)^a\widehat{T}_{23}(z)\frac{T_{12}(\bu)|0\rangle}{\hat\lambda_2(z)\lambda_1(\bu) } .
\end{equation}
Substituting here $\widehat{T}_{23}(z)$ from \eqref{HT} we find
\begin{equation}\label{BVap11}
   \widehat{\mathbb{B}}_{0,a+1}(\emptyset;\{\bu,z\})  =
(-1)^a\bigl(T_{13}(z)T_{32}(z-c) - T_{12}(z)T_{33}(z-c)\bigr)\frac{T_{12}(\bu)|0\rangle}{\hat\lambda_2(z)\lambda_1(\bu)} .
\end{equation}
To calculate the obtained action we use commutation relations of the monodromy matrix entries. The $RTT$-relation
\eqref{RTT} implies
\be{rtt}
[T_{ij}(u),T_{kl}(v)]=g(u,v)\bigl( T_{kj}(v)T_{il}(u)-T_{kj}(u)T_{il}(v)\bigr).
\ee
In particular, we have
\be{rtt1}
T_{32}(u)T_{12}(v)= T_{12}(v)T_{32}(u)f(u,v)-T_{12}(u)T_{32}(v)g(u,v).
\ee
We see that permuting the operators $T_{32}$ and $T_{12}$ we obtain the annihilation operator $T_{32}$
on the right. Eventually, this operator approaches the vector $|0\rangle$ and  annihilates it.
Thus, the contribution from the term $T_{13}(z)T_{32}(z-c)$ vanishes.

The commutation relations \eqref{rtt} also imply
\be{rtt2}
T_{33}(u)T_{12}(v)= T_{12}(v)T_{33}(u)+g(u,v)\bigl( T_{13}(v)T_{32}(u)-T_{13}(u)T_{32}(v)\bigr).
\ee
We see that when the operator $T_{33}$ is permuted with the operator $T_{12}$, it either commutes
or generates the operator $T_{32}$. As we have already seen, the latter annihilates the state $T_{12}(\bu)|0\rangle$.
Thus, the operator $T_{33}(z-c)$ acts on the state $T_{12}(\bu)|0\rangle$ as
\be{actT33}
T_{33}(z-c)T_{12}(\bu)|0\rangle=\lambda_{3}(z-c)T_{12}(\bu)|0\rangle.
\ee
Substituting this into \eqref{BVap11}, we arrive at
\begin{equation}\label{BVap12}
   \widehat{\mathbb{B}}_{0,a+1}(\emptyset;\{\bu,z\})  =
(-1)^{a+1}\lambda_{3}(z-c)\frac{T_{12}(z)T_{12}(\bu)|0\rangle}{\lambda_1(\bu)\hat\lambda_2(z)}
=(-1)^{a+1} \frac{\lambda_2(\bu)\lambda_2(z)}{\lambda_1(\bu)\lambda_1(z) }   \mathbb{B}_{a+1,0}(\{\bu,z\};\emptyset),
\end{equation}
what completes the first step of the induction. Thus, equation \eqref{hB-B} holds for $b=0$ and $a$ arbitrary non-negative.

\subsection{Second step of induction\label{S-SSI}}

We pass to the second step of induction.  This time we use a recursion for the Bethe vectors $\widehat{\mathbb{B}}$ \cite{BelPRS12c}
\begin{multline}\label{Recu2}
\hat\lambda_2(z)g(\bu,z)\widehat{\mathbb{B}}_{b+1,a}(\{\bv+c,z\};\bu)=\widehat{T}_{12}(z)\widehat{\mathbb{B}}_{b,a}(\bv+c;\bu)\\
+\sum_{j=1}^a g(u_j,z)\frac{f(\bu_j,u_j)}{g(u_j,\bv)}\widehat{T}_{13}(z)\widehat{\mathbb{B}}_{b,a-1}(\bv+c;\bu_j).
\end{multline}
This recursion allows us to uniquely construct the Bethe vector $\widehat{\mathbb{B}}_{b+1,a}$, knowing the
Bethe vectors\footnote{We set by definition $\widehat{\mathbb{B}}_{b,-1}=0$.}
$\widehat{\mathbb{B}}_{b,a}$ and $\widehat{\mathbb{B}}_{b,a-1}$.

Assume that \eqref{hB-B} holds for some $b\ge 0$ and $a$ arbitrary.
Then we can replace the Bethe vectors $\widehat{\mathbb{B}}$ by $\mathbb{B}$ in the rhs of
\eqref{Recu2}. We obtain
\begin{multline}\label{RecuhB-b}
\hat\lambda_2(z)g(\bu,z)\widehat{\mathbb{B}}_{b+1,a}(\{\bv+c,z\};\bu)=
(-1)^{a+b+ab}\frac{\lambda_2(\bu)\lambda_2(\bv)}{\lambda_1(\bu)\lambda_3(\bv)}\Biggl\{
\widehat{T}_{12}(z)\;\mathbb{B}_{a,b}(\bu;\bv)\\
+(-1)^{b+1}\sum_{j=1}^a \frac{\lambda_1(u_j)g(u_j,z)f(\bu_j,u_j)}{\lambda_2(u_j)g(u_j,\bv)}\widehat{T}_{13}(z)
\;\mathbb{B}_{a-1,b}(\bu_j;\bv)\Biggr\}.
\end{multline}
We should compute the action of the operator
$\widehat{T}_{12}(z)$ on $\mathbb{B}_{a,b}(\bu;\bv)$ and the action of the operator
$\widehat{T}_{13}(z)$ on $\mathbb{B}_{a-1,b}(\bu_j;\bv)$. This is done in sections~\ref{S-A12} and~\ref{S-A13} respectively.
The results have the following form:
\be{hT13BVi}
\widehat{T}_{13}(z)\mathbb{B}_{a-1,b}(\bu_j;\bv)=(-1)^{a+b}\lambda_{2}(z)\lambda_{2}(z-c)\frac{g(z,\bv)}{h(\bu_j,z)}
\mathbb{B}_{a,b+1}(\{\bu_j,z\};\{\bv,z-c\}),
\ee
and
\begin{multline}\label{acthT12}
\widehat{T}_{12}(z)\mathbb{B}_{a,b}(\bu;\bv)=(-1)^{a+1}\lambda_{2}(z-c)\Bigr\{\lambda_{1}(z)g(\bu,z)\mathbb{B}_{a,b+1}(\bu;\{\bv,z-c\})\\
+\lambda_{2}(z)g(\bv,z)\sum_{j=1}^a \frac{\lambda_1(u_j) g(z,u_j)f(\bu_j,u_j)}
{\lambda_2(u_j)g(\bv,u_j)h(\bu_j,z) }\mathbb{B}_{a,b+1}(\{\bu_j,z\};\{\bv,z-c\})\Bigr\}.
\end{multline}
Substituting these formulas into \eqref{RecuhB-b} we immediately arrive at
\begin{equation}\label{Recuhsub}
\hat\lambda_2(z)\widehat{\mathbb{B}}_{b+1,a}(\{\bv+c,z\};\bu)=
(-1)^{1+b+ab}\frac{\lambda_2(\bu)\lambda_2(\bv)}{\lambda_1(\bu)\lambda_3(\bv)}\lambda_{1}(z)\lambda_{2}(z-c)
\mathbb{B}_{a,b+1}(\bu;\{\bv,z-c\}).
\end{equation}
Finally, using $\hat\lambda_2(z)=\lambda_{1}(z)\lambda_{3}(z-c)$ we obtain
\begin{equation}\label{Indstep}
\widehat{\mathbb{B}}_{b+1,a}(\{\bv+c,z\};\bu)=
(-1)^{a+(b+1)+a(b+1)}\frac{\lambda_2(\bu)\lambda_2(\bv)\lambda_{2}(z-c)}{\lambda_1(\bu)\lambda_3(\bv)\lambda_{3}(z-c)}
\mathbb{B}_{a,b+1}(\bu;\{\bv,z-c\}).
\end{equation}
This completes the second step of the induction.

\section{Action formulas \label{S-AF}}

\subsection{Actions of the operators $T_{ij}$ on Bethe vectors $\mathbb{B}_{a,b}$\label{S-ATB}}

In this section we give a list of formulas for the actions of the operators $T_{ij}(z)$ on the Bethe vectors $\mathbb{B}_{a,b}(\bu;\bv)$.
These formulas were obtained in \cite{BelPRS12c}. Here they are adopted to the new normalization of the Bethe vectors.
In all action formulas $\bar\eta=\{z,\bu\}$ and $\bar\xi=\{z,\bv\}$. We also set
\be{Lamb}
\Lambda(z)=\frac{\lambda_2(z)}{h(\bv,z)h(z,\bu)}.
\ee
\begin{itemize}

\item Action of $T_{13}(z)$:
 \be{act13}
 T_{13}(z)\mathbb{B}_{a,b}(\bu;\bv)=\Lambda(z)\mathbb{B}_{a+1,b+1}(\bet;\bxi).
 \ee

\item Action of $T_{12}(z)$:
 \be{act12}
 T_{12}(z)\mathbb{B}_{a,b}(\bu;\bv)=\Lambda(z)\sum_{\#\bxi_{\so}=1}
 \frac{f(\bxi_{\st},\bxi_{\so})h(\bxi_{\so},\bet)}{h(z,\bxi_{\so})}\,
 \mathbb{B}_{a+1,b}(\bet;\bxi_{\st}).
 \ee
The sum is taken over partitions  $\bxi\Rightarrow\{\bxi_{\so},\bxi_{\st}\}$ so that $\#\bxi_{\so}=1$.

\item Action of $T_{23}(z)$:
 \be{act23}
 T_{23}(z)\mathbb{B}_{a,b}(\bu;\bv)=\Lambda(z)\sum_{\#\bet_{\so}=1}
\frac{f(\bet_{\so},\bet_{\st})h(\bxi,\bet_{\so})}{h(\bet_{\so},z)}\,
 \mathbb{B}_{a,b+1}(\bet_{\st};\bxi).
 \ee
The sum is taken over partitions  $\bet\Rightarrow\{\bet_{\so},\bet_{\st}\}$
so that $\#\bet_{\so}=1$.


\item Action of $T_{22}(z)$:
 \be{act22}
 T_{22}(z)\mathbb{B}_{a,b}(\bu;\bv)=\Lambda(z)\sum_{\#\bxi_{\so}=\#\bet_{\so}=1}
 \frac{f(\bet_{\so},\bet_{\st})f(\bxi_{\st},\bxi_{\so})h(\bxi_{\so},\bet)h(\bxi_{\st},\bet_{\so})}
 {h(z,\bxi_{\so})h(\bet_{\so},z)}\,
 \mathbb{B}_{a,b}(\bet_{\st};\bxi_{\st}).
 \ee
The sum is taken over partitions  $\bxi\Rightarrow\{\bxi_{\so},\bxi_{\st}\}$
and $\bet\Rightarrow\{\bet_{\so},\bet_{\st}\}$ so that $\#\bxi_{\so}=\#\bet_{\so}=1$.

\item Action of $T_{11}(z)$:
 \be{act11}
 T_{11}(z)\mathbb{B}_{a,b}(\bu;\bv)=\Lambda(z)\sum_{\#\bxi_{\so}=\#\bet_{\so}=1}
 \frac{\lambda_1(\bet_{\so})f(\bet_{\st},\bet_{\so})f(\bxi_{\st},\bxi_{\so})h(\bxi_{\so},\bet_{\st})}
 {\lambda_2(\bet_{\so})g(\bxi_{\st},\bet_{\so})h(z,\bxi_{\so})}\,
 \mathbb{B}_{a,b}(\bet_{\st};\bxi_{\st}).
 \ee
The sum is taken over partitions  $\bxi\Rightarrow\{\bxi_{\so},\bxi_{\st}\}$
and $\bet\Rightarrow\{\bet_{\so},\bet_{\st}\}$ so that $\#\bxi_{\so}=\#\bet_{\so}=1$.

\item Action of $T_{21}(z)$:
 \begin{multline}\label{act21}
 T_{21}(z)\mathbb{B}_{a-1,b}(\bu;\bv)=\Lambda(z)\sum_{\substack{\#\bet_{\so}=\#\bet_{\st}=1\\ \#\bxi_{\so}=1}}
 \frac{\lambda_1(\bet_{\so})f(\bet_{\st},\bet_{\so})f(\bet_{\st},\bet_{\sth})f(\bet_{\sth},\bet_{\so})f(\bxi_{\st},\bxi_{\so})}
 {\lambda_2(\bet_{\so})g(\bxi_{\st},\bet_{\so})h(\bet_{\st},z)   h(z,\bxi_{\so})}\\
\times  h(\bxi_{\so},\bet_{\st})h(\bxi_{\so},\bet_{\sth})h(\bxi_{\st},\bet_{\st})\mathbb{B}_{a-1,b}(\bet_{\sth};\bxi_{\st}).
 \end{multline}
The sum is taken over partitions  $\bxi\Rightarrow\{\bxi_{\so},\bxi_{\st}\}$
and $\bet\Rightarrow\{\bet_{\so},\bet_{\st},\bet_{\sth}\}$ so that
$\#\bxi_{\so}=\#\bet_{\so}=\#\bet_{\st}=1$.

\end{itemize}
The actions of $\widehat{T}_{ij}$ onto $\widehat{\mathbb{B}}_{a,b}(\bu;\bv)$ are given by the same formulas,
where we should put hats for the operators, the vacuum eigenvalues $\lambda_k(z)$, and the Bethe vectors.

\subsection{Actions of the operators $\widehat{T}_{ij}$ on Bethe vectors $\mathbb{B}_{a,b}$\label{S-AhTB}}

The action formulas \eqref{act13}--\eqref{act21} allow us to derive the actions of the operators
$\widehat{T}_{ij}$ onto the Bethe vectors $\mathbb{B}_{a,b}(\bu;\bv)$. For this we should
express $\widehat{T}_{ij}$ in terms of the original entries $T_{ij}$ via \eqref{SklAldop}--\eqref{Qmin}.
In particular, we have
\begin{equation}\label{HT}
\begin{aligned}
 & \widehat{T}_{12}(z) = T_{21}(z)T_{13}(z-c) - T_{23}(z)T_{11}(z-c),\\
&  \widehat{T}_{23}(z) = T_{13}(z)T_{32}(z-c) - T_{12}(z)T_{33}(z-c),\\
&  \widehat{T}_{13}(z) = T_{12}(z)T_{23}(z-c) - T_{13}(z)T_{22}(z-c).
\end{aligned}
\end{equation}
Then the actions of $\widehat{T}_{ij}$ onto  $\mathbb{B}_{a,b}(\bu;\bv)$ can be obtained via
successive application of the formulas \eqref{act13}--\eqref{act21}. Below
we give some details of this derivation for the action of $\widehat{T}_{13}$ and $\widehat{T}_{12}$.


\subsubsection{Action of $\widehat{T}_{13}$\label{S-A13}}

The operator $\hat T_{13}(z)$ is given by the last equation \eqref{HT}. It is convenient to consider the following
combination
\be{combi}
T_{12}(x)T_{23}(y) - T_{13}(x)T_{22}(y)
\ee
and set $x=z$, $y=z-c$ in the end. Such the replacement of $\widehat{T}_{13}(z)$ allows us to avoid singular expressions in
the intermediate computations.

Applying successively, first \eqref{act22} and \eqref{act13}, and  then \eqref{act23} and \eqref{act12} we obtain
\begin{equation}\label{T13T22}
T_{13}(x)T_{22}(y)\mathbb{B}_{a,b}(\bu;\bv)=\frac{\Lambda(x,y)}{h(y,x)}\sum_{\substack{\#\bxi_{\so}=1\\ \#\bet_{\so}=1}}
\frac{f(\bet_{\so},\bet_{\st})f(\bxi_{\st},\bxi_{\so})\;
h(\bxi_{\so},\bet)h(\bxi_{\st},\bet_{\so})}
{h(y,\bxi_{\so})h(\bet_{\so},y) f(x,\bxi_{\so})f(\bet_{\so},x)}\mathbb{B}_{a+1,b+1}(\bet_{\st};\bxi_{\st}),
\end{equation}
and
\begin{equation}\label{T12T23}
T_{12}(x)T_{23}(y)\mathbb{B}_{a,b}(\bu;\bv)=\frac{\Lambda(x,y)}{h(y,x)}
\sum_{\substack{\#\bxi_{\so}=1\\ \#\bet_{\so}=1}}
\frac{f(\bet_{\so},\bet_{\st})f(\bxi_{\st},\bxi_{\so})h(\bxi_{\so},\bet)h(\bxi_{\st},\bet_{\so})}
{h(\bet_{\so},y) h(x,\bxi_{\so})f(\bet_{\so},x)}\mathbb{B}_{a+1,b+1}(\bet_{\st};\bxi_{\st}).
\end{equation}
Here $\bet=\{\bu,x,y\}$ and $\bxi=\{\bv,x,y\}$.  The sum is taken over partitions
$\bet\Rightarrow\{\bet_{\so},\bet_{\st}\}$ and $\bxi\Rightarrow\{\bxi_{\so},\bxi_{\st}\}$ so that
$\#\bet_{\so}=\#\bxi_{\so}=1$. Here we also introduced
\be{Lxy}
\Lambda(x,y)=\frac{\lambda_{2}(x)\lambda_{2}(y)}{h(x,y)h(\bv,x)h(\bv,y)h(x,\bu)h(y,\bu)}.
\ee

Taking the difference of \eqref{T13T22} and \eqref{T12T23} we arrive at
\be{HT13}
\widehat{T}_{13}(z)\mathbb{B}_{a,b}(\bu;\bv)=\Lambda(x,y)
\sum_{\substack{\#\bxi_{\so}=1\\ \#\bet_{\so}=1}}
\frac{f(\bet_{\so},\bet_{\st})f(\bxi_{\st},\bxi_{\so})h(\bxi_{\so},\bet)h(\bxi_{\st},\bet_{\so})}
{h(\bet_{\so},y) h(x,\bxi_{\so})h(y,\bxi_{\so})f(\bet_{\so},x)}\mathbb{B}_{a+1,b+1}(\bet_{\st};\bxi_{\st})\Bigr|_{\substack{x=z\\y=z-c}}\;.
\ee

Now we should consider several cases. First of all, we see that $\bet_{\so}\ne x$, because otherwise  the factor $1/f(\bet_{\so},x)$ in
\eqref{HT13} is equal to zero.
Thus, either $\bet_{\so}=y$ or $\bet_{\so}=u_j$, $j=1,\dots,a$. Consider the first case $\bet_{\so}=y$ and denote this contribution by $G$. Then
\be{G1}
G=\Lambda(x,y)f(y,\bu)h(\bv,y)h(x,y)
\sum_{\#\bxi_{\so}=1}\frac{f(\bxi_{\st},\bxi_{\so})h(\bxi_{\so},x)h(\bxi_{\so},\bu)}
{h(x,\bxi_{\so})h(y,\bxi_{\so})}\mathbb{B}_{a+1,b+1}(\{\bu,x\};\bxi_{\st})\Bigr|_{\substack{x=z\\y=z-c}}\;.
\ee
This case respectively should be divided into subcases.
\begin{itemize}
\item $\bxi_{\so}=x$, hence, $\bxi_{\st}=\{\bv,y\}$. Then, substituting \eqref{Lxy} in \eqref{G1} we find
\be{G11}
G_1=(-1)^{a+b+1}\lambda_{2}(z)\lambda_{2}(z-c)\frac{g(z,\bv)}{h(\bu,z)}
\mathbb{B}_{a+1,b+1}(\{\bu,z\};\{\bv,z-c\}).
\ee

\item $\bxi_{\so}=y$, hence, $\bxi_{\st}=\{\bv,x\}$. Then $h(\bxi_{\so},x)=h(y,x)\to 0$, as $x\to z$ and $y\to z-c$. Thus,
this contribution vanishes.

\item $\bxi_{\so}=v_j$, $j=1,\dots,b$, hence, $\bxi_{\st}=\{\bv_j,x,y\}$. Then
\be{G13}
G_1^{(2)}=
\frac{\lambda_{2}(x)\lambda_{2}(y)g(y,\bu)f(\bv_j,v_j)g(x,v_j)g(y,v_j)h(v_j,\bu)}
{h(\bv_j,x)h(x,\bu)}
\mathbb{B}_{a+1,b+1}(\{\bu,x\};\{\bv_j,x,y\})\Bigr|_{\substack{x=z\\y=z-c}}\;.
\ee
In this case the Bethe vector $\mathbb{B}_{a,b}(\{\bu,x\};\{\bv_j,x,y\})$ vanishes in the limit $x=z$ and $y=z-c$.
Indeed, we have due to \eqref{act13}
 \be{act13inv}
 \mathbb{B}_{a+1,b+1}(\{\bu,x\};\{\bv_j,x,y\}) = \frac1{\lambda_2(x)}h(y,x)h(\bv_j,x)h(x,\bu)  T_{13}(x)\mathbb{B}_{a,b}(\bu;\{\bv_j,y\}),
 \ee
and the rhs of \eqref{act13inv} vanishes, because $h(z-c,z)=0$.

\end{itemize}

Similarly, one should consider the case $\bet_{\so}=u_j$, $j=1,\dots,a$. The analysis of this case shows that all
the corresponding contributions vanish. Thus, the action of $\widehat{T}_{13}(z)$ on the Bethe vector $\mathbb{B}_{a,b}(\bu;\bv)$ is
given by \eqref{G11}:
\be{hT13BV}
\widehat{T}_{13}(z)\mathbb{B}_{a,b}(\bu;\bv)=(-1)^{a+b+1}\lambda_{2}(z)\lambda_{2}(z-c)\frac{g(z,\bv)}{h(\bu,z)}
\mathbb{B}_{a+1,b+1}(\{\bu,z\};\{\bv,z-c\}).
\ee

\subsubsection{Action of $\widehat{T}_{12}$\label{S-A12}}

The action of $\widehat{T}_{12}(z)$ can be considered exactly in the same manner. Using \eqref{HT} and the action
formulas \eqref{act13}--\eqref{act21} we obtain
\begin{multline}\label{hT12}
\widehat{T}_{12}(z)\mathbb{B}_{a,b}(\bu;\bv)=\Lambda(x,y)
\sum_{\substack{\#\bet_{\so}=\#\bet_{\st}=1\\ \#\bxi_{\so}=1}}
\frac{\lambda_1(\bet_{\so})f(\bet_{\st},\bet_{\so})f(\bet_{\st},\bet_{\sth})f(\bet_{\sth},\bet_{\so})
f(\bxi_{\st},\bxi_{\so})}
{\lambda_2(\bet_{\so})g(\bxi_{\st},\bet_{\so})h(\bet_{\st},x) h(x,\bxi_{\so})h(y,\bxi_{\so})}\\
\times h(\bxi,\bet_{\st})h(\bxi_{\so},\bet_{\sth})\mathbb{B}_{a,b+1}(\bet_{\sth};\bxi_{\st})
\Bigr|_{\substack{x=z\\y=z-c}}\;.
\end{multline}
Here $\bet=\{\bu,x,y\}$ and $\bxi=\{\bv,x,y\}$. The sum is taken over partitions
$\bet\Rightarrow\{\bet_{\so},\bet_{\st},\bet_{\sth}\}$ and $\bxi\Rightarrow\{\bxi_{\so},\bxi_{\st}\}$ so that
 $\#\bet_{\so}=\#\bet_{\st}=\#\bxi_{\so}=1$.

Again one should consider several cases. The analysis shows that non-vanishing contributions arise if and only if $\bxi_{\so}=x$
and $\bet_{\st}=y$. Then
\begin{multline}\label{hT12nv}
\widehat{T}_{12}(z)\mathbb{B}_{a,b}(\bu;\bv)=\frac{\lambda_{2}(x)\lambda_{2}(y)g(y,\bu)g(\bv,x)}{h(x,\bu)}\\
\times \sum_{\#\bet_{\so}=1}
\frac{\lambda_1(\bet_{\so})f(\bet_{\sth},\bet_{\so})h(x,\bet_{\sth})}
{\lambda_2(\bet_{\so})g(\bv,\bet_{\so})g(y,\bet_{\so})}
\mathbb{B}_{a,b+1}(\bet_{\sth};\{\bv,y\})
\Bigr|_{\substack{x=z\\y=z-c}}\;,
\end{multline}
where $\bet=\{\bu,x\}$ and the sum is taken over partitions
$\bet\Rightarrow\{\bet_{\so},\bet_{\sth}\}$  so that  $\#\bet_{\so}=1$. Then we should consider two cases.
First, we can set $\bet_{\so}=x$ and $\bet_{\sth}=\bu$. Then we obtain the first term in \eqref{acthT12}.
The second case is $\bet_{\so}=u_j$ and $\bet_{\sth}=\{\bu_j,x\}$, $j=1,\dots,a$.
Then we obtain the second term in \eqref{acthT12}.

\end{document}